\documentclass[12pt]{article}
\usepackage{latexsym}
\usepackage{amstext}
\usepackage{amsmath}
\usepackage{mathtools}
\usepackage[normalem]{ulem}
\usepackage{subfigure}
\usepackage{epsfig, graphicx}
\usepackage{color}
\usepackage{amsfonts,bm,color}
\usepackage{verbatim}
\usepackage{float}
\usepackage{latexsym}
\usepackage{setspace}
\newcommand{\ds}{\displaystyle}
\newcommand{\beq}{\begin{eqnarray}}
\newcommand{\eeq}{\end{eqnarray}}
\newcommand{\beqq}{\begin{eqnarray*}}
\newcommand{\eeqq}{\end{eqnarray*}}
\newcommand{\p}{\partial}

\newcommand{\eps}{\varepsilon}

\newcommand{\x}{\mbox{\boldmath$x$}}

\newcommand{\s}{\mbox{\boldmath$s$}}

\begin{document}
%%%%%%%%%%%%%%%%%%%%%%%%%%%%%%%%%%%%%%%%%%%%%%%%%%%%%%
\title{Geometrical effects on nonlinear electrodiffusion in cell physiology}
 %\author{A. Singer}
%\email{amits@math.princeton.edu} \affiliation{Department of
%Mathematics and PACM, Princeton University, Fine Hall, Washington
%Road, Princeton NJ 08544-1000 USA}
\author{J. Cartailler$^{1}$,  Z. Schuss\footnote{Department of Mathematics, Tel-Aviv University,
Tel-Aviv 69978, Israel.}, and D. Holcman\footnote{ $^1$ Ecole Normale
Sup\'erieure, 46 rue d'Ulm 75005 Paris, France and Mathematical Institute, University of Oxford,
Andrew Wiles Building, Woodstock Rd, Oxford OX2 6GG, United Kingdom. \hspace {3 cm} {Corresponding
author email:david.holcman@ens.fr}}}
\date{\today}
%%%%%%%%%%%%%%%%%%%%%%%%%%%%%%%%%%%%%%%%%%%%%%%%%%%%%%

\maketitle
%\doublespacing
\begin{abstract}
We report here new electrical laws, derived from nonlinear electro-diffusion theory, about the effect of the local geometrical structure, such as curvature, on the electrical properties of a cell.  We adopt the Poisson-Nernst-Planck (PNP) equations for charge concentration and electric potential as a model of electro-diffusion. In the case at hand, the entire boundary is impermeable to ions and the electric field satisfies the compatibility condition of Poisson's equation. We construct an asymptotic approximation for certain singular limits to the steady-state solution in a ball with an attached cusp-shaped funnel on its surface. As the number of charge increases, they concentrate at the end of cusp-shaped funnel.  These results can be used in the design of nano-pipettes and help to understand the local voltage changes inside dendrites and axons with heterogenous local geometry.
\end{abstract}
%\pacs{05.40.-Jc., 87.10.-e}
%We further study the case of charges in a narrow ellipse and inside a corrugated channel.In all cases, the density of charges accumulates near points of local maximum
%curvature of the boundary.

%%%%%%%%%%%%%%%%%%%%%%%%%%%%%%%%%%%%%%%%%%%%%%%%%%%%%%
\section{Introduction}
%%%%%%%%%%%%%%%%%%%%%%%%%%%%%%%%%%%%%%%%%%%%%%%%%%%%%%
Electro-diffusion is the process by which the motion of ions in solution is driven by two physical forces: thermal motion, which is diffusion, and the electric field. The difficulty in the mathematical description of this physical motion is due to the origin of the field, which consists of the contribution of mobile ions and of a possible external field. The dielectric membrane also affects the field by image charges. So far only few electro-diffusion systems are well understood: although the voltaic cell was invented more than 200 years ago, designing optimal configurations is still a challenge. On the other extreme, ionic flux and gating of voltage-channels \cite{Bezanilla} is now well explained by the modern Poisson-Nernst-Planck theory of electro-diffusion \cite{Nitzan}, because at the nanometer scale, the cylindrical geometry approximation of protein channels reduces the computation of the electric field and of ionic diffusion to one dimension \cite{Eisenberg1,Roux,Roux2,Eisenberg2,EKS,nadlerschuss,Singer}. However, cellular domains at a micron scale involve two- and three-dimensional geometry, much more complicated than the cylindrical geometry of a channel pore, leading to a more complex electro-diffusion description \cite{Savtchenko,HY2015}.

We recall that local curvature is a key geometrical element for controlling charge distribution in various media, such as in the air (e.g., the lightning rod \cite{CH2}). The manifestation of this effect is observed in Lebesgue's thorn, which is a an inverted cusp singularity of the boundary, for which the solution of Laplace's equation blows-up inside the domain \cite[p.304]{CH2}. In electronics, the design of printed circuits is always pre-conditioned on corner effects \cite{Ruiz}. However, these effects are not very well known inside an electrolytic bath. Recent analysis \cite{PhysD2016}, \cite{HY2015} suggests that non-electro-neutrality in the geometry of an electrolyte confined by a dielectric membrane affects charge distribution.

{We use the Poisson-Nernst-Planck (PNP) equations for charge concentration and electric potential as a model of electro-diffusion. The entire boundary is impermeable to particles (ions) and the electric field satisfies the compatibility condition of Poisson's equation. Phenomenological descriptions of electro-diffusion, such as the cable equation or the reduced electrical engineering approximation by resistance, capacitance, and even electronic devices, are not sufficient to describe non-cylindrical geometry \cite{HY2015}, because they assume a simple reduced one-dimensional or reduced geometry. We present here results about charge and field distributions  in electro-diffusion in various geometrical microdomains, when the condition of electro-neutrality is not satisfied.}  We recall that under the non-electro-neutrality assumption, and with charge distributed in bounded domains
confined by a dielectric membrane, Debye's concept of charge screening decaying exponentially away from a charge \cite{Debye}, do not apply and long-range correlation leads to a gradient of charges in a ball with no inward current. A new capacitance law was derived for an electrolyte ball \cite{PhysD2016}, where the difference of potential between the center $C$ and the surface $S$, that is, $V(C)-V(S)$, increases, first linearly and then logarithmically, when the total number of charges in the ball increases.

Our aim here is to understand the effect of boundary curvature on an electrical cell,
such as neuron. In particular, we explore the effect of boundary curvature on the charge and field distribution at steady state. The curvature of membranes of dendrites and axons of neurons have many local maxima that can modulate the channel's local electric potential \cite{YusteBook}. In this  article, we study the effects of local curvature on the distribution of charge in bounded domains with no electro-neutrality. The effect of non-electro-neutrality was recently studied in \cite{PhysD2016} and a long-range electrostatic length, much longer than the Debye length was found. This effect is due to the combined effects of non-electro-neutrality and boundary, which lead to charge accumulation near the boundary.

The cusp-shaped funnel geometry was studied in \cite{Direstrait2012}, however this paper presents several crucial mathematical differences with \cite{Direstrait2012}, in particular, we are solving a nonlinear equation, while it was linear in \cite{Direstrait2012}. Furthermore, the boundary condition at the end of the cusp-shaped funnel: while it is the Dirichlet condition in \cite{Direstrait2012}, it is the Neumann condition here. This means that in \cite{Direstrait2012} the absorption flux at the end of the funnel is computed, whereas here the stationary voltage and charge distribution are computed in the absence of flux. We develop here new boundary layer analysis, different than the classical matched asymptotics method \cite{Ward1,Ward2,Ward3}. The manuscript is organized as follow: first, we consider a bounded domain with an uncharged narrow cusp-shaped funnel on the boundary, which is a singular geometrical effect. Second, we further study the case of charge distribution in a charged narrow cusp.

%%%%%%%%%%%%%%%%%%%%%%%%%%%%%%%%%%%%%%%%%%%%%%%%%%%%%%
\section{The PNP equations}\label{s:PNP-eq}
%%%%%%%%%%%%%%%%%%%%%%%%%%%%%%%%%%%%%%%%%%%%%%%%%%%%%%
The Poisson-Nernst-Planck system of equations in a domain $\Omega$, whose
dielectric boundary $\p\Omega$ is represented as the compatibility condition for
Poisson's equation, and its impermeability to the passage of ions is represented as a
no-flux boundary condition for the Nernst-Planck equation. We assume that the total charge in $\Omega$ consists of $N$ identical positive ions with initial particle density $q(\x)$ in $\Omega$, their valence is $z$, and the total number of particles is fixed, equal to
\begin{align}
\int\limits_\Omega q(\x)\,d\x=N.
\end{align}
Thus the charge in $\Omega$ is
\beqq
Q=zeN,
\eeqq
where $e$ is the electronic charge. The charge density $\rho(\x,t)$ is the solution of the
{initial and boundary value problem for the} Nernst-Planck equation
\begin{align}
D\left[\Delta \rho(\x,t) +\frac{ze}{kT} \nabla \left(\rho(\x,t) \nabla \phi(\x,t)\right)\right]=&\,
\frac{\p\rho(\x,t)}{\p t}\hspace{0.5em}\mbox{for}\ \x\in\Omega\label{NPE}\\
D\left[\frac{\p\rho(\x,t)}{\p n}+\frac{ze}{kT}\rho(\x,t)\frac{\p\phi(\x,t)}{\p
n}\right]=&\,0\hspace{0.5em}\mbox{for}\ \x\in\p\Omega \label{noflux}\\
\rho(\x,0)=&\,q(\x)\hspace{0.5em}\mbox{for}\ \x\in\Omega.\label{IC}
\end{align}
Here $\phi(\x,t)$ is the electric potential in $\Omega$ and is the solution of the Neumann problem for the Poisson equation
\begin{align}
\label{poisson} \Delta \phi(\x,t) =&\,
-\frac{ze\rho(\x,t)}{\eps\eps_0}\hspace{0.5em}\mbox{for}\ \x\in\Omega\\
\frac{\p\phi(\x,t)}{\p
n}=&\,-\sigma(\x,t)\hspace{0.5em}\mbox{for}\ \x\in{\p\Omega},\label{Boundary_Phi} % -\frac{\sigma(\x,t)}{\eps\eps_0}
\end{align}
where $\sigma(\x,t)$ is the surface charge density on the boundary $\p\Omega$. In the steady state,
\begin{align}
\sigma(\x,t)=\frac{Q}{\eps\eps_0 |\p \Omega|}.% +\frac{Q}{4\pi R^2}
\end{align}
 %%%%%%%%%%%%%%%%%%%%%%%%%%%%%%%%%%%%%%%%%%%%%%%%%%%%%%
\section{Steady solution in a ball with a cusp-shaped funnel}\label{ss:SSS}
\begin{figure}[http!]
	\center
	\includegraphics[scale=0.65]{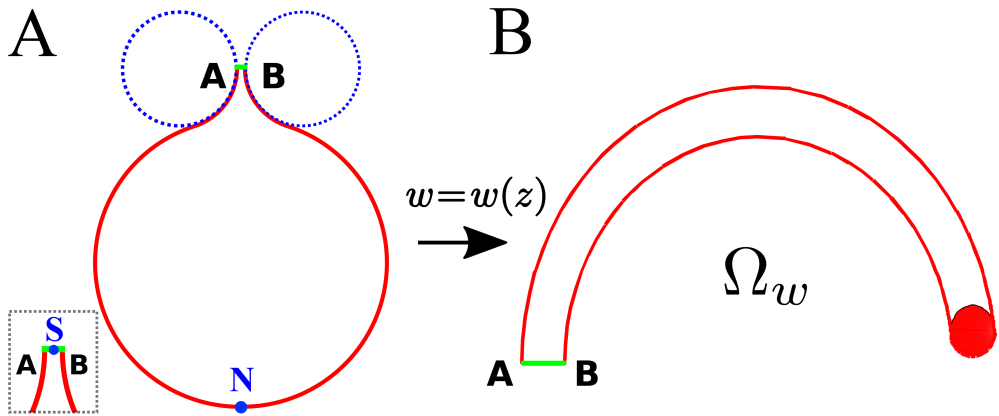}
%\centering \resizebox{!}{6cm}{\includegraphics{domainsCusp2}}
%\centering \resizebox{!}{5cm}{\includegraphics{domainsCuspConf-2}}
\caption[Conformal image of a funnel]{\setstretch{1.5} \small Image $\Omega_w=w(\Omega)$ of
the domain $\Omega$ ({\bf A.}) under the conformal mapping \eqref{w}. The neck (left) is mapped onto the
semi-annulus enclosed between the like-style arcs and the large disk in $\Omega$ is mapped
onto the small red disk. The short green segment $AB$ (left) (of length $\eps$) is
mapped onto the thick green segment $AB$ (of length $2\sqrt{\eps}+O(\eps)$). The letters $S$ and $N$ designate the south and the north pole respectively.} \label{f:conf}
\end{figure}

%%%%%%%%%%%%%%%%%%%%%%%%%%%%%%%%%%%%%%%%%%%%%%%%%%%%%%
{Local boundary curvature is a key geometrical feature that controls charge distribution
in the domain. Specifically, we study the effect of a narrow funnel attached to a sphere}. In
various media, such as air (e.g., the lightning rod, \cite{CH2}), the
manifestation of this effect is observed in Lebesgue's thorn, which is a an inverted cusp
singularity of the boundary, for which the solution of Laplace's equation blows-up inside the domain
\cite[p.304]{CH2}.
In the steady state  \eqref{NPE} gives the particle density
\begin{align}
\rho(\x)=N\frac{\exp\left\{-\ds\frac{ze\phi(\x)}{kT}\right\}}
{\ds\int_\Omega\exp\left\{-\ds\frac{ze\phi(\x)}{kT}\right\}\,d\x},\label{N}
\end{align}
hence \eqref{poisson} gives {Poisson equation}
\begin{align}
\Delta\phi(\x)=-\frac{zeN\exp\left\{-\ds\frac{ze\phi(\x)}{kT}\right\}}{\eps\eps_0{\ds\int_\Omega
\exp\left\{-\ds\frac{ze\phi(\x)}{kT}\right\}\,d\x}}{.}\label{Deltaphi}
\end{align}
and \eqref{Boundary_Phi} gives the boundary condition
\beq
\frac{\p\phi({\x})}{\p n}=-\frac{Q}{\eps\eps_0 |\p \Omega|}, \label{compatibility}
\eeq
{for $|\x|=R$}, which is the compatibility condition, obtained by integrating Poisson's equation \eqref{poisson} over $\Omega$.
%The inequality in \eqref{eqsymm1} means that $\phi(r)$ has a maximum at the origin and decreases %toward the boundary (see Fig.~\ref{f:LambdaS}A).
 Changing variables to
\beq \label{conversion}
u(\x)=\ds \frac{ze \phi({\x})}{kT},\quad\lambda= \frac{(ze)^2N}{\eps\eps_0 kT},
\eeq
Poisson's equation \eqref{Deltaphi} becomes
\begin{align}
\Delta u(\x)=&\, -\frac{\lambda \exp \left\{-\ds u(\x) \right\}}{\ds\int_{\Omega} \exp \left\{-\ds
u(\x)\right\}\, d\x}\label{eqsymm}%\\
%u(0)=&\,0,\quad u'(0)=0.\nonumber
\end{align}
and the boundary condition \eqref{compatibility} becomes
\beq
\frac{\p u(\x)}{\p n}=-\frac{\lambda}{|\p \Omega|}\hspace{0.5em} \mbox{for}\ \x\in\p
\Omega.
\eeq
%To study the effect of the cusp, we use the Mobius transformation
%First, we consider the case $\nu_\pm=1$, $\ell_+=R_c$, and $l_-=r_c$, radius
%$1$, and $A$ has dimensionless radius $r_c/R_c$. This case can represent a
%partial block described in Figure \ref{f:Partial-block}(left).
%Under the scaling \eqref{scaling}
The translation $\tilde{u}=u+{\ds\ln\left(\lambda/\int_\Omega\exp\{v(\x)\}\,d\x \right)}$, converts \eqref{eqsymm}  into
 \begin{align}\label{NewPb}
-\Delta \tilde{u}(\x)=&\,\exp\{-\tilde{u}(\x)\}\hspace{0.5em}\mbox{for}\ \x\in\Omega\\
\frac{\p \tilde{u}(\x)}{\p n}=&\,-\frac{\lambda}{|\p \Omega|}\hspace{0.5em}\mbox{for} \ \x\in\p\Omega.\nonumber
\end{align}
{We} consider a dimensionless {planar} domain {$\Omega$} with a cusp-shaped
funnel formed by two bounding circles $A$ and $B$ of dimensionless radii 1 (see Fig.\ref{f:conf}(left)). The opening of the funnel is $\eps\ll1$. We construct an asymptotic solution {in this limit} to  the nonlinear boundary value problem (BVP) \eqref{NewPb} by first mapping the domain $\Omega$ conformally {with}  the M\"obius transformation of the two {osculating} circles $A$ and $B$ into concentric circles (see {Fig.}\ref{f:conf}(right)). To this end, we move the origin of the complex plane to the center of the {osculating} circle $B$ and set
\beq\label{w}
w=w(z)=\frac{z-\alpha}{1-\alpha z},
\eeq
where
\beq\label{alpha1}
\alpha=-1-\sqrt{\eps}+O(\eps).
\eeq
The M\"obius transformation \eqref{w} maps the circle $B$ (dashed blue) into itself and $\Omega$ is mapped onto the domain $\Omega_w=w(\Omega)$ in Figure \ref{f:conf}{(right)}.
%%%%%%%%%%%%%%%%%%%%%%%%%%%%%%%%%%%%%%%%%%%%%%%%%%%%%%%%%%%%%%%%%%%%%%%%%%5
%%%%%%%%%%%%%%%%%%%%%%%%%%%%%%%%%%%%%%%%%%%%%%%%%%%%%%%%%%%%%%%%%%%%%%%5
The straits in Figure \ref{f:conf}(left) are mapped onto the ring enclosed between the like-style
arcs and the large disk is mapped onto the small {red} disk in Figure \ref{f:conf}(right). The
radius of the small disk and the elevation of its center above the real axis are $O(\sqrt{\eps})$.
The short black segment $AB$ of length $\eps$ in Figure \ref{f:conf}(left) is mapped onto the
segment $AB$ of length $2\sqrt{\eps}+O(\eps)$ in Figure \ref{f:conf}(right). This mapping
(see \cite{hoze2011}), transforms the PNP equations as well and thus leads
to a new non-linear effect.
Setting $u(z)=v(w)$ converts \eqref{eqsymm} to
\begin{align}
\ds \Delta_w v(w)=&\, \ds-\frac{\exp \left\{-\ds v(w) \right\}}{|w'(z)|^2}\nonumber\\
=&\,-\ds \frac{(4\eps+O(\tilde\eps^{3/2}))
}{|w(1-\sqrt{\tilde\eps})-1+O(\tilde\eps)|^4} \exp \left\{-\ds v(w) \right\}\hspace{0.5em}
\mbox{ for } w \in\Omega_w.\label{PDEw}
\end{align}
The boundary {segment} $AB$ at the end {of the cusp-shaped funnel} in {Figure}
\ref{f:conf}(left) is denoted $\p\Omega_{w,a}$. To determine the boundary conditions,
we use the change of coordinates $w=Re^{i\theta}=X+iY$. At the end {of the funnel}, where
$R  \simeq 1$, we get
\beq
\frac{\p u(z)}{\p n_{z}}={\left.-\frac{\p v(w)}{\p \theta}\right|_{{w=-1}}\frac{\p \theta}{\p Y},}
\eeq
where
\beq
i e^{i\theta}\frac{\p \theta}{\p Y}= w'(z)=\frac{1-\alpha^2}{(1-\alpha z)^2}{.}
\eeq
For $\theta=\pi$ (for $z=-1$), we obtain {$\p \theta/\p Y= -2/\sqrt{\eps}$} and the boundary condition at $\p\Omega_{w,a}$ is
\beq
\ds {\frac{\p v(w)}{\p n}=-\frac{\lambda \sqrt{\eps}}{2|\p \Omega |}\hspace{0.5em}\mbox{for}\
w\in\p\Omega_{w,a}.}\label{BC}
\eeq
%%%%%%%%%%%%%%%%%%%%%%%%%%%%%%%%%%%%%%%%%%%%%%%%%%
\subsection{Reduced PNP equations in an uncharged cusp-shaped funnel}
%%%%%%%%%%%%%%%%%%%%%%%%%%%%%%%%%%%%%%%%%%%%%%%%%%
%%%%%%%%%%%%%%%%%%%%%%%%%%%%%%%%%%
\begin{figure}[http!]
	\center
	\includegraphics[scale=0.64]{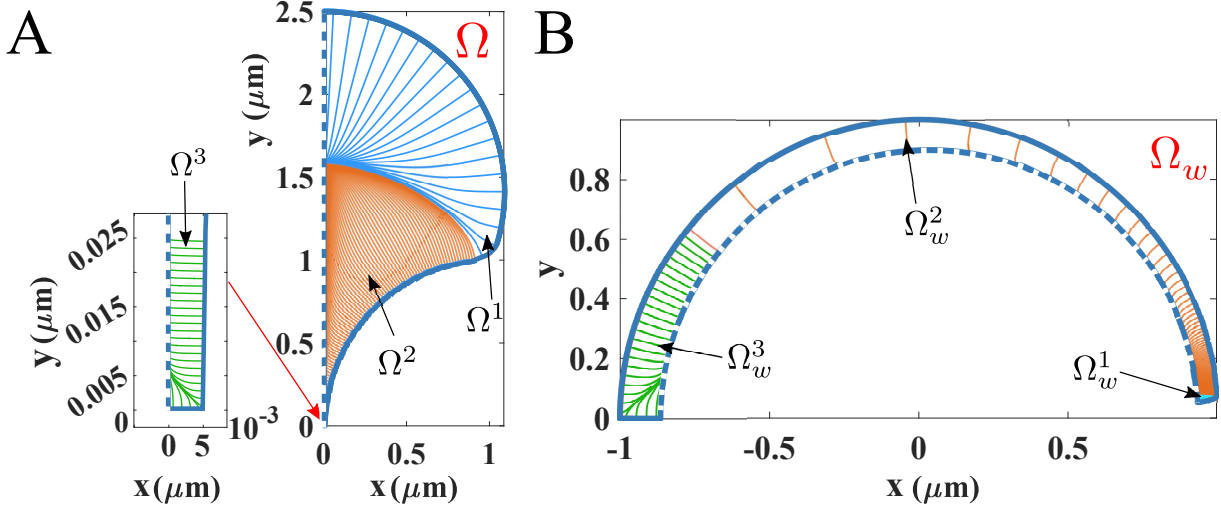}
	\caption{\small \setstretch{1.5} {\bf Influence of the cusp on the field lines (orthogonal to the level lines)}. The field line inside the original domain $\Omega$ {\bf (A)} and its image domain $\Omega_{w}$ {(\bf B),}  computed numerically from equation \eqref{NewPb}. The blue lines originate from the bulk, while the orange starts in the cusp. The domain $\Omega_{w}$ is subdivided into three regions: the region $\Omega^1_{w}$ inside the funnel, the region $\Omega^2_{w}$ connecting the end of the funnel to the bulk $\Omega^3_{w}$.}
\label{f:Field}
\end{figure}
\begin{figure}[http!]
 	\center
 	\includegraphics[scale=0.22]{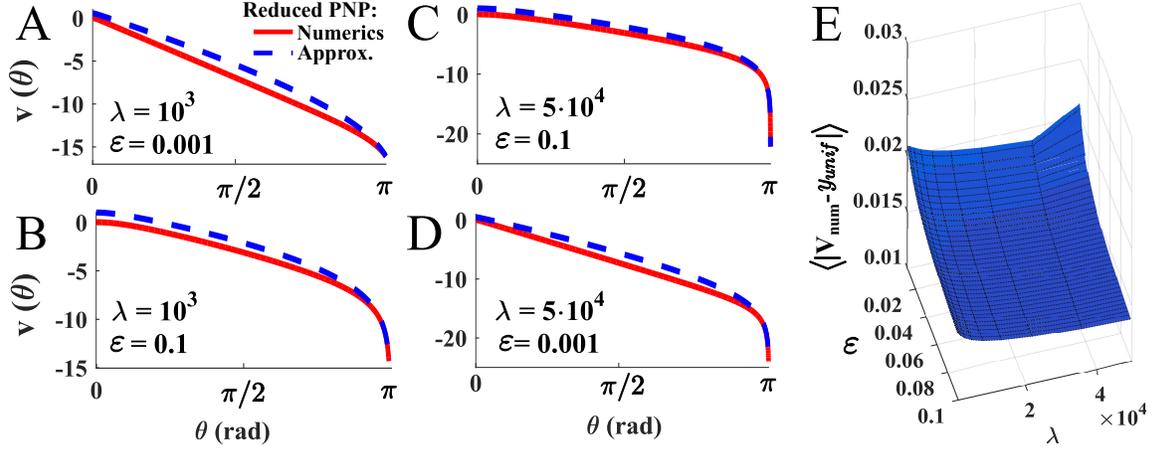}
 	\caption {\setstretch{1.5}\small{{{The asymptotic solution  $y_{\mbox{\scriptsize unif}}(\theta)$  of \eqref{unif} (blue dashed lines) is compared to the numerical solution of \eqref{reduced_1} (red line).  The four panels {\bf A-B-C-D}  are obtained for different pairs of parameters $(\lambda, \eps)$.}}}  {\bf E.} shows a 3D plots the difference between the asymptotic solution $y_{\mbox{\scriptsize unif}}$ (eq. \ref{unif}) and numerical results $V_{num}$ (eq. \ref{reduced_1}), averaged over the domain $\Omega_w$}.
 	\label{f:Sol_NoCharge}
 \end{figure}

%%%%%%%%%%%%%%%%%%%%%%%%%%%%%%%%%%
Approximating the banana-shaped domain $\Omega_{w}$ by a one-dimensional circular arc, we use a one-dimensional approximation of the solution in  $\Omega_{w}$ \cite{HS2012,HS2015}. This approximation assumes
that there are no {non-neutralized charges} on the surface of the cusp (Fig.\ref{f:Sol_NoCharge}A). The boundary condition for the approximate  one-dimensional solution of \eqref{PDEw} is zero at
angle $\theta_{Lim}=c\sqrt{\eps}$, where $c$ is a constant (see details in \cite{HS2012,HS2015}) and
represents the solution inside the disk in Figure \ref{f:conf}(left), away from the
cusp. Thus,  \eqref{PDEw} in the conformal image $\Omega_{w}$ becomes the boundary value problem
\begin{align}
v''+\frac{4\eps}{|e^{i\theta}-1-e^{i\theta}\sqrt{\eps}|^4}\exp \left\{-\ds v(e^{i\theta})
\right\}=&\,0\label{reduced_1}\\
v'(c\sqrt{\eps})=&\,0\label{req}\\
v'(\pi)=&\,-\frac{\lambda \sqrt{\eps}}{2|\p \Omega |}.\nonumber
\end{align}
The solution of \eqref{reduced_1} is shown in Figure \ref{f:Sol_NoCharge}B-C in the two domains, $\Omega$ {(panel {\bf A}) and its image $\Omega_w$ (panel {\bf B})}.
%%%%%%%%%%%%%%%%%%%%%%%%%%%%%%%%%%
%\begin{figure}[http!]
%\centering \resizebox{!}{8cm}{\includegraphics{cuspsolution}}
%\caption{\small General solution in a cusp} \label{f:solution}
%\end{figure}
%%%%%%%%%%%%%%%%%%%%%%%%%%%%%%%%%%%%%%%%5

Our goal is now to estimate the difference of potentials between the north pole $N$ and the end of the funnel $C$,
\beq\label{diffp}
\tilde \Delta u = u(N)-u(C)=v(c\sqrt{\eps})-v(\pi){.}
\eeq
To {construct an asymptotic approximation to the solution of \eqref{req} in the limits
$\eps\to0$ and $\lambda\to\infty$, we first construct the outer-solution in the form of a {series in
powers of $\eps$}, which is an approximation valid away from the boundary. In the limit of small
$\eps$, the first term in the series vanishes,} exponential terms drop out, and the second order
term is
\beq
y_{\mbox{\scriptsize outer}}(\theta)=M\theta+M',
\eeq
where $M$ and $M'$ are yet undetermined constants. The outer solution cannot satisfy all boundary conditions, so a boundary layer correction is needed at the reflecting boundary at $\theta=c\sqrt{\eps}$. Thus, we set $\theta=\sqrt{\eps}\xi$
and expand
\beqq
\frac{\eps^2}{|e^{i\theta}-1-e^{i\theta}\sqrt{\eps}|^4}
=\frac{1}{(1+\xi^2)^2}+O(\sqrt{\eps}).
\eeqq
%\beqq
%\frac{\eps^2}{|e^{i\eta}-1-e^{i\eta}\sqrt{\eps}|^4}
%=\frac{1}{(1+\xi^2)^2}+O(\sqrt{\eps}).
%\eeqq
Writing the boundary layer solution as $y_{\mbox{\scriptsize bl}}(\theta)=Y(\xi)$, we obtain to
leading order the boundary layer equation
\beq
Y''(\xi)+\frac{4}{(1+\xi^2)^2}\exp{\left\{-Y(\xi)\right\}}=0,\label{BLEQ}
\eeq
%\beq
%Y''(\xi)+\frac{1}{4(1+\xi^2)^2}\exp{\left\{-Y(\xi)\right\}}=0,\label{BLEQ}
%\eeq
with $Y'(c)=0$. The solution is decaying  for large $\xi$ and develops a singularity at finite $\xi$. However, a Taylor expansion near $\xi=0$,
\beq
Y(\xi)=A+B_2\xi^2+B_4\xi^4+\ldots\label{Yxi},
\eeq
gives in \eqref{BLEQ}
\beq
{B_2}=-2e^{-A}.
\eeq
%\beq
%{B_2}=-\frac{e^{-A}}{8}.
%\eeq
In general, the coefficients satisfy $B_k=O(e^{-A})$, for $A\gg1$. For small $\xi$, we
obtain the approximate solution of \eqref{BLEQ} by {considering the leading term in a
regular expansion of the solution in powers of  $\xi$. The equation for the leading term is
\beq
Y''(\xi)+\frac{4e^{-A}}{(1+\xi^2)^2}=0\label{BLEQa}
\eeq
and the solution is defined up to an additive constant. Setting  $Y_{appr}(0)=0$, which does not affect the
potential difference, we find that
\beq
Y_{appr}(\xi)=-2\xi e^{-A}\arctan \xi.\label{BLEQa2}
\eeq
It follows that the boundary layer solution at $c\sqrt{\eps}$ is
\beq
y_{\mbox{\scriptsize bl}}(\theta)=A-\frac{2\theta}{\sqrt{\eps}} e^{-A}\arctan{\frac{\theta}{\sqrt{\eps}}}.
\eeq
The boundary layer near $\pi$  {is needed, because $A\to\infty$ as $\eps\to0$ (see \eqref{Atoinfty} below).}
An approximation of the solution can be obtained by freezing the power-law term in \eqref{req}, for
which the  equation is for a generic parameter $b>0$,
\begin{align*}
\frac{d^2 }{d \theta^2} v(\theta) +b e^{-v(\theta)}=0,\quad\frac{d v(0)}{d \theta}= v(0)=0.
\end{align*}
The solution is
\beq \label{blowups}
v_b(\theta)={\ln\cos^2\frac{b}{2}\theta.}
\eeq
Putting the outer and boundary layer solutions together gives the uniform asymptotic approximation
\beq\label{unif}
{y_{\mbox{\scriptsize unif}}(\theta)=A-\frac{2\theta}{ \sqrt{\eps}} e^{-A}\arctan
\frac{\theta}{\sqrt{\eps}}+\ln\cos^2\frac{b}{2}\theta,}
\eeq
where the parameters $A$ and $b$ are yet undetermined constants. The condition at
{$c\sqrt{\eps}=o(1)$ for $\eps\ll1$} is satisfied, because
\beqq
y_{\mbox{\scriptsize unif}}'(0)=0.
\eeqq
The condition at {$\theta=\pi$} gives that
\beq\label{cond}
{y_{\mbox{\scriptsize unif}}'(\pi)=-\frac{\pi e^{-A}}{ \sqrt{\eps}} -b\tan\frac{b}{2}\pi= \ds-\frac{\lambda
\sqrt{\eps}}{2|\p \Omega |}.\nonumber}
\eeq
The compatibility condition for  \eqref{NewPb},
\beq\label{Compatibility2new}
\lambda= \int\limits_{\Omega}\exp \{-\tilde u(\x)\} dS_{\x},
\eeq
gives in {$\Omega_w$} that
\beq\label{Compatibility2}
\lambda = \int\limits_{{\Omega_w}}\exp \{-\tilde v(w)\} \frac{dw}{|\phi'(\phi^{-1}(w))|}=  8\sqrt{\eps}\int\limits_{c\sqrt{\eps}}^\pi \frac{ \exp{\{-v(\theta)\}}}{| e^{i\theta}(1-\sqrt{\eps}) - 1  |^4 }{\, d\theta.}
\eeq
Using the uniform approximation \eqref{unif} in the compatibility condition  \eqref{Compatibility2}, we obtain the second condition
\begin{align} \label{Compatibility3}
\lambda =&\, 8\sqrt{\eps}e^{-A}\int\limits_{c\sqrt{\eps}}^\pi \frac{1}{\cos^2\ds\frac{b}{2}\theta}\frac{{\exp\left\{\ds
e^{-A}\frac{2\theta}{ \sqrt{\eps}} \arctan\frac{\theta}{\sqrt{\eps}}\right\}}}{| e^{i\theta}(1-\sqrt{\eps}) - 1  |^4 }\,
d\theta\nonumber \\
\approx&\, \frac{8e^{-A}}{\eps}\int\limits_{0}^{\pi/\sqrt{\eps}} \frac{1}{\cos^2\ds\frac{b}{2}\sqrt{\eps}\xi}\frac{
\exp{\left\{\ds2e^{-A} \xi\arctan \xi\right\}}}{|1+\xi^2  |^2 }\,d\xi,
\end{align}
where we used the change of variable $\theta=\sqrt{\eps}\xi$. Integrating by parts, we get {for $\eps\ll1$}
\beq\label{Compatibility3b}
\lambda\sim \frac{8e^{-A}}{\eps} \left( \frac{2}{b\sqrt{\eps}}\tan\frac{b}{2}\pi\frac{ \exp{\left\{\ds 2e^{-A} \frac{\pi}{\sqrt{\eps}}\frac{\pi}{2}\right\}}}{\left|1+\left(\ds\frac{\pi}{\sqrt{\eps}}\right)^2
\right|^2 }-\int\limits_{0}^{\pi/\sqrt{\eps}} \frac{2}{b\sqrt{\eps}}\tan\frac{b}{2}\theta\ {\Psi(\theta)\,d\theta}\right),
\eeq
where
\beq
\Psi(\xi)={
\frac{d}{d\xi}\frac{ \exp\left\{\ds  2 e^{-A} \xi\arctan \xi\right\}}{|1+\xi^2  |^2 }}.
\eeq
Thus, it remains to solve the asymptotic equation
\beq\label{Compatibility3b1}
\lambda\sim 8e^{-A}\eps^{1/2} \left[ \frac{2}{b\pi^4}\tan\frac{\pi  b}{2}\exp{\left\{\ds \frac{\pi^2e^{-A}}{ \sqrt{\eps}}\right\}}+ O\left(\ln\left|{\cos\ds\frac{\pi b}{2}}\right|\right)\right].
\eeq
for $A$ and $b$ in the limit $\eps\to0$.
We consider the limiting case where
\beq\label{Case_1}
\frac{e^{-A}}{\sqrt{\eps}} ={O(1)=C\hspace{0.5em}\mbox{for}\ \lambda\to\infty,}
\eeq
for which condition {\eqref{cond}} can be simplified and gives to leading order
\beq
b\tan{\frac{\pi b}{2}}=\frac{\lambda \sqrt{\eps}}{2|\p \Omega |}\label{btan},
\eeq
that is, for $\lambda\sqrt{\eps}\ll1$ \eqref{btan} gives
$$b\approx 1-\frac{4}{\pi} \frac{|\p \Omega |}{\lambda\sqrt{\eps}},\quad
\tan{\frac{b}{2}\pi\sim} \frac{\lambda\sqrt{\eps}}{2|\p \Omega |}.$$
With condition \eqref{Compatibility3b1}, we get
\beq\label{Compatibility3b2}
\lambda\approx 8e^{-A}\eps^{1/2} \left[\frac{2}{\pi^4}\frac{\lambda\sqrt{\eps}}{2|\p \Omega |} \exp\left\{\ds \frac{\pi^2e^{-A}}{ \sqrt{\eps}}\right\}+ O\left(\ln\left|\cos{\frac{\pi b}{2}}\right|\right)\right],
\eeq
To leading order in large $C$, we obtain
\beq
\frac{\pi^4 |\p\Omega|}{8\eps^{3/2}}= C \exp{ \left\{ C\pi^2 \right\}}.
\eeq
The solution is expressed in terms of the Lambert-W function,
\beq
C\pi^2=W\left(\frac{\pi^6 |\p\Omega|}{2^3\eps^{3/2}}\right),
\eeq
and for small $\eps$, using the asymptotics of the Lambert function,
\beq
C\pi^2 =\ln\frac{\pi^6 |\p\Omega|}{2^3\eps^{3/2}}-\ln{\left[\ln\frac{\pi^6
|\p\Omega|}{2^3\eps^{3/2}}\right]}+o(1).
\eeq
Finally,
\begin{align}
\frac{e^{-A}}{\sqrt{\eps}}=&\, C\sim\frac{1}{\pi^2}\ln\frac{\pi^6
|\p\Omega|}{2^3\eps^{3/2}},\nonumber\\
A=&\,\ln\frac{1}{\sqrt{\eps}}- \ln{\left[\frac{1}{\pi^2}\ln\frac{\pi^6
|\p\Omega|}{2^3\eps^{3/2}}\right]}{\to\infty\hspace{0.5em}\mbox{as}\ \eps\to0.}
\label{Atoinfty}
\end{align}
It follows that a uniform asymptotic approximation \eqref{unif} in the limits $\lambda\to\infty$ %and
$\eps\to0$ %{such that $\lambda\eps=O(1)$ {\bf IS THIS CONDITON ASSUMED?},
is given by
\begin{align}\label{unif2}
y_{\mbox{\scriptsize unif}}(\theta)=&\,\ln\frac{1}{\sqrt{\eps}}-
\ln{\left[\frac{1}{\pi^2}\ln\frac{\pi^6 |\p\Omega|}{2^3\eps^{3/2}}\right]}\\
&\,-2{\theta} \frac{1}{\pi^2}\ln\frac{\pi^6 |\p\Omega|}{2^3\eps^{3/2}}\arctan
\frac{\theta}{\sqrt{\eps}}+\ln{\left[\cos^2\frac{1-\frac{4}{\pi} \frac{|\p \Omega|}{\lambda\sqrt{\eps}}}{2}\theta \right]}.\nonumber
\end{align}
%\begin{align}\label{unif2}
%y_{\mbox{\scriptsize unif}}(\theta)=&\,\ln\frac{1}{\sqrt{\eps}}-
%\ln{\left[\frac{16}{\pi^2}\ln\frac{\pi^6 |\p\Omega|}{2^7\eps^{3/2}}\right]}\\
%&\,-2{\theta} \frac{1}{\pi^2}\ln\frac{\pi^6 |\p\Omega|}{2^7\eps^{3/2}}\arctan
%\frac{\theta}{\sqrt{\eps}}+\ln{\left[\cos^2\frac{1-\frac{4}{\pi} \frac{|\p \Omega|}{\lambda\sqrt{\eps}}}{2}\theta \right]}.\nonumber
%\end{align}
The uniform approximation} \eqref{unif2} is plotted for different values of $\eps$ and $\lambda$ in Figure \ref{f:Sol_NoCharge} against the numerical {solution} of \eqref{reduced_1}, with the boundary conditions
$v'(c\sqrt{\eps})=v'(0)=0$. The numerical solutions are computed with the software COMSOL, based on an adaptive mesh refinement and a relative tolerance of $10^{-3}$, that we validated on known analytical results of steady state PNP equations in a disk \cite{PhysD2016}. We find that the asymptotic expansion is particularly good in the
limit $\eps\rightarrow0$ and $\lambda\rightarrow \infty$ (Fig.\ref{f:Sol_NoCharge}A-D). However,
for $\lambda=O(1)$ the log-term approximation in  \eqref{unif2}  is non-monotonic in $\theta$.
%%%%%%%%%%%%%%%%%%%%%%%%%%%%%%%%%%%%%%%%%%%%%%%%%%%
%%%%%%%%%%%%%%%%%%%%%%%%%%%%%%%%%%%%%%%%%%%%%%%%%%%%%
Finally, to further validate the uniform asymptotic expansion, we compared the numerical solutions
of the full {equation \eqref{diffp}} in the initial domain $\Omega$ with the reduced PNP
{equation \eqref{NewPb}} with zero Neumann boundary conditions, except at
the end of the funnel for the mapped domain $\Omega_w$. The result is shown in Figure \ref{f:Sol_NoCharge}A-D, showing good agreement between the one-dimensional PNP approximation in $\Omega_w$ and the numerical solution of the full equation. We estimated numerically the difference between the asymptotic solution $y_{\mbox{\scriptsize unif}}$ \eqref{unif} and the numerical estimation $V_{num}$ \eqref{reduced_1}, averaged over the domain $\Omega_w$, for $10^3\leq \lambda \leq 5\cdot 10^4 $ and $5\cdot10^{-3} \leq \eps \leq 10^{-1}$. The difference is almost constant in the range $[0.01,0.025]$.

To compare the voltage at the north and south poles (at the end of the funnel), we use the two-dimensional analytical solution in the
entire ball and the numerical solution of \eqref{NewPb} (Fig.\ref{f:Sol_NoCharge}D). Interestingly, we find that the difference
$u(N)-u(S)$ has a maximum with respect to $\lambda$, where $u(N)$  and $u(S)$ are the values of the potential at the north pole and at the end of the funnel, respectively.
%%%%%%%%%%%%%%%%%%%%%%%%%%%%%%%%
\subsection{The voltage drop between the end of the funnel and the center of the ball}\label{ss:ld}
%%%%%%%%%%%%%%%%%%%%%%%%%%%%%%%%%
We can now use \eqref{unif} to compute the potential drop in \eqref{diffp}. It is given by
\begin{align}\label{diffpp3}
\tilde \Delta_{S C} u= &\,u(S)-u(C)=-v(c\sqrt{\eps})+v(\pi)\nonumber\\
=&\,-\ln\frac{\pi^6 |\p\Omega|}{2^3\eps^{3/2}}+2\ln\frac{2|\p\Omega|}{\lambda\eps^{1/2}}
=\ln\frac{2^5|\p\Omega|\sqrt{\eps}}{\pi^6 \lambda^2}.
\end{align}

The potential difference $\tilde \Delta_{S C} u$ with respect to $\lambda$ is shown in Figure
\ref{f:Sol_NoCharge}F (red line).

{Next, we compare the potential drop \eqref{diffp}} with the one between the center and the north pole. Numerical {solution of the PNP equations} shows that the voltage and charge distribution in a disk with a funnel do not differ from the ones {in} a disk in the upper sphere {(Fig.\ref{f:Field}). This result is compared next to} the difference between the north pole and the center {evaluated} from the exact analytical expression derived for a disk.

The expression for the voltage in the two-dimensional disk of radius $R$ is given by {(see \cite{PhysD2016})}
\beqq\label{Solution_Disc}
u_{\lambda}^{2D}(x)=\ln {\left[1-\frac{\lambda_D}{8\pi+\lambda_D} \left(\frac{r}{R}\right)^2\right]^2,}
\eeqq
where $\lambda_D$ is a parameter. We calibrate $\lambda_D$ so that the solutions of the PNP equations in a disk with a funnel have the same total charge as a disk. The Neumann boundary conditions for the disk and the funnel are, respectively,
\beqq
\frac{\p {u}(\x)}{\p n}=- \ds\frac{\lambda_D}{2\pi R},\quad
\ds \frac{\p {u}(\x)}{\p n}=\ds -\frac{\lambda}{|\p \Omega|}.
\eeqq
The calibration is
\beq\label{lambda_D}
\ds \lambda_D=\lambda \,\frac{2\pi R }{|\p \Omega|}.
\eeq
We compare in Figure \ref{f:Sol_NoCharge}D the two-dimensional numerical solution of the PNP
equation \eqref{NewPb} in the domain $\Omega$ (blue line), with the analytical solution
\eqref{Solution_Disc} in a disk with no cusp (dashed red). The numerical solution of the PNP
equation \eqref{NewPb} is plotted along the main axis $0y$ in the interval $[0,y_0]$ (where the
point $y_0$ is defined by the condition $\nabla u(y_0)=0$). In the range $[y_0,y_{cusp}]$, where
$y_{cusp}$ is the coordinate of the cusp, we compare the solution of \eqref{NewPb} with the uniform
solution $y_{unif}$ of \eqref{unif} in the funnel (dashed green). We conclude that in the cusp, the two-dimensional approximation in a disk is in good agreement with the
numerical solution of equation \eqref{NewPb}, confirming that the solution in the bulky head does not influence the one in the cusp (as already shown in Fig. \ref{f:Field}). This result
also confirms the validity of the analytical formula to predict the large $\lambda$ asymptotics.

For a disk of radius $R$, the potential drop is given by
\beq\label{diffpp2dr1}
\tilde \Delta_{NC} u = u(N)-u(C)=\ln \left(\frac{8\pi}{8\pi+\lambda_D}\right)^2= -2\ln \lambda -2\ln\left(\frac{R}{4|\p \tilde \Omega|}\right)+O(\frac{1}{\lambda})
\eeq
(see section \ref{ss:ld}). The {potential drop} $\tilde \Delta_{NC} u$ is shown in Figure \ref{f:Sol_NoCharge}E (blue line). The two differences of potential $\bar \Delta_{SC} u$ \eqref{diffpp2dr1} and $\tilde \Delta_{N C} u$ \eqref{diffpp3} have the same logarithmic behavior $\ln{1/\lambda^2}$ for $\lambda\gg1$ and $u(N)-u(S)=O(1)$. A numerical solution  in two-dimensions shows that $u(N)-u(S)$ may converge to zero as
$\lambda$ increases (Fig. \ref{f:Sol_NoCharge}F), thus having a local maximum for small values of $\lambda$. This maximum cannot be analyzed by the uniform expression \eqref{unif}, because it appears outside the domain of validity of \eqref{unif}. This result is in
agreement with the two-dimensional numerical solution of \eqref{NewPb} for the difference between $u(N)$ (potential at the north pole) and $u(S)$ (potential at the end of the  funnel) (Fig. \ref{f:Sol_NoCharge}F).
%%%%%%%%%%%%%%%%%%%%%%%%%%%%%%%%%%%%%%%%%%%%%%%%%%%%%%%%%%%%%%%%%%%%%%%%%%%%%%%5
\begin{figure}[http!]
	\center
	\includegraphics[scale=0.8]{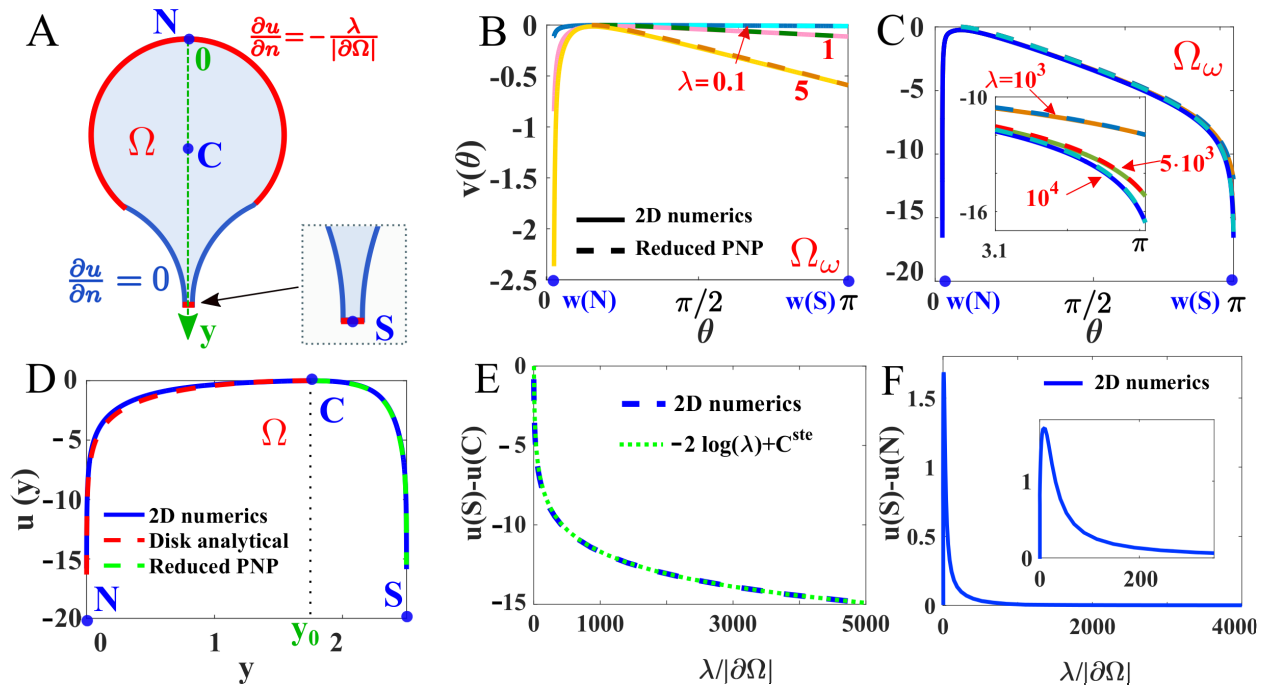}
	\caption {\small\setstretch{1.5} {Comparison of the numerical solutions of the full and reduced PNP
equations \eqref{NewPb} with zero Neumann boundary conditions, except at the end of the funnel.}
{\bf A.} Schematic representation of the domain $\Omega$ with an {uncharged} cusp (blue).
The letters $N$, $S$, and $C$ refer to the north pole, the funnel tip, and the center of mass respectively.
{\bf B-C} Numerical solutions of \eqref{NewPb} (solid) and the solution of \eqref{PDEw4} in the funnel (dashed) in the mapped domain  $\Omega_w$. The solution have been obtained for $\eps=0.01$.
{\bf D.} Comparison
of \eqref{NewPb} (blue) with the numerical solution \eqref{reduced_1} inside the funnel (dashed
green) and \eqref{Solution_Disc} in the bulk (dashed red).
{\bf E.} Solution $u(S)-u(C)$ (dashed blue) obtained numerically from \eqref{diffpp3} and compared to the logarithmic function $-2\ln(\lambda)$ (greed dotted).
{\bf F.} Two-dimensional numerical solutions of the difference $|u(N)-u(C)|$ vs $\lambda$. The inset in panel {\bf F.} is a blowup showing a maximum for small $\lambda$. }  \label{f:Sol_NoCharge2}
\end{figure}
%%%%%%%%%%%%%%%%%%%%%%%%%%%%%%%%%%%%%%%%%%%%%%%%%%%%%%%%%%%%%
{The potential drop calculated above is non-dimensionalized by the radius of curvatures
 $R_f$  at the right and left of the funnel,
\beqq
 \eps = \frac{\tilde \eps}{R_f} ,
\eeqq
where {$\tilde \eps$ is the length of the absorbing arc $AB$. The non-dimensionalized
volume and boundary measure are, respectively,
\[|\Omega|=\frac{|\tilde \Omega|}{R^2},\quad |\p\Omega|=\frac{|\p\tilde \Omega|}{R}.\]

In dimensional units \eqref{diffpp3} gives the potential drop in the dimensional disk with a funnel as}
\beq\label{diffpp3dim}
\tilde \Delta_{S C} u = u(S)-u(C)=\ln{\frac{2^5|\p \tilde \Omega|\sqrt{\tilde \eps}}{\pi^6 R_f^{3/2}\lambda^2}}.
\eeq
We conclude  in the limit of $\lambda\gg1$,  $\tilde\eps\to0$ that the difference of potential between the end of cusp $S$ and the north pole $N$ in the domain is obtained by adding \eqref{diffpp2dr1} and \eqref{diffpp3dim} and we get
\beq\label{diffpp3dimRf}
\tilde \Delta_{S N} u = u(S)-u(C)+u(C)-u(N)=\ln\left(\frac{2^5 |\p \tilde \Omega|\sqrt{\tilde \eps}}{\pi^6 {R_f^{3/2}}}\right) +2\ln\left(\frac{R}{4|\p \tilde \Omega|}\right)+O(\frac1{\lambda}).
\eeq
We recall that $R$ is the radius of the entire ball, while $R_f$ is the radius of curvature of the funnel.
%%%%%%%%%%%%%%%%%%%%%%%%%%%%%%%%%%%%%%%%%%%%%%%%%%
\section{The PNP equations in a charged domain with a cusp-shaped funnel}\label{s:cc}
%%%%%%%%%%%%%%%%%%%%%%%%%%%%%%%%%%%%%%%%%%%%%%%%%%
Due to the Neumann boundary condition \eqref{Boundary_Phi} on the lateral part of the
funnel,\eqref{PDEw} in the transformed domain cannot be reduced to one dimension. Thus
we derive a different one-dimensional approximation for the mapped PNP equations in the banana-shaped domain $\Omega_w$ by averaging over the radius $r$. Rewriting  \eqref{PDEw} in polar coordinates $w=re^{i\theta}$, we obtain
\beq
\ds\frac{1}{r}\frac{\p}{\p r }\left( r\frac{\p v(w) }{\p r} \right) +\frac{1}{r^2}\frac{\p^2 v(w)
}{\p \theta^2} =\,-\ds \frac{(4\eps+O(\eps^{3/2}))\,\exp \left\{-\ds v(w) \right\}
}{|re^{i\theta}(1-\sqrt{\eps})-1+O(\eps)|^4} \hspace{0.5em} \mbox{ for } w\in\Omega_w.\label{PDEw2}
\eeq
In the section $\Omega_w\cap \{1-\sqrt{2\eps}<r<1]\}$, the boundary conditions are
\begin{align}\label{bcondi}
\left.\frac{\p v(r,\theta)}{\p r }\right|_{r=1}=&\,\frac{-\lambda\sqrt{\eps}}{|\p
\Omega|(\cos\theta-1)}, \, \hbox{ for } \theta \in [c\sqrt{\eps}, \pi]\\
\left.\frac{\p v(r,\theta)}{\p r }\right|_{r=1-\sqrt{2\eps}}=&\, 0, \, \hbox{ for } \theta \in
[c\sqrt{\eps}, \pi]\nonumber\\
\left.\frac{\p v(r,\theta)}{\p \theta }\right|_{ \theta=\pi}=&\,\frac{-\lambda\sqrt{\eps}}{2|\p
\Omega|}, \nonumber\\
\left. \frac{\p v(r,\theta)}{\p \theta }\right|_{\theta=c\sqrt{\eps}}=&\, 0.\nonumber
%\left.\frac{\p v(r,\theta)}{\p  \theta }\right|_{\theta=c\sqrt{\eps}}=&\, 0.\nonumber\\
\end{align}
%where $C$ is a constant that should be chosen to match the bulk potential.
{Taylor's expansion of $v$ in the section  gives
\beq\label{expand}
v(r,\theta)=v_0(\theta)+(r-1)v_1(\theta)+O((r-1)^2),
\eeq
{and because} $|r-1|=O(\sqrt{\eps})$, we obtain the  approximation,
\beqq
\exp \left\{-\ds v(w) \right\}=\exp \left\{-\ds v_0(\theta) \right\}\left(1- \sqrt{\eps} v_1(\theta) +O(\eps)\right ).
\eeqq
Multiplying \eqref{PDEw2} by $r^2$ and} integrating over the radius, we get
\beq
\left[ r\frac{\p v(r,\theta) }{\p r} \right]_{1-\sqrt{\eps}}^{r=1}
+{\frac{\p^2}{\p \theta^2}\int\limits^{1}_{1-\sqrt{\eps}}v(r,\theta)\, dr=}\,-
\int\limits_{1-\sqrt{\eps}}^1\frac{(4r^{2}\eps+O(\eps^{3/2}))\,e^{-\ds v(r,\theta)} }{|re^{i\theta}(1-\sqrt{\eps})-1+O(\eps)|^4} \,dr. \label{pdeW2}
\eeq
The boundary conditions \eqref{bcondi} give, to leading order in $\sqrt{\eps}$, {that}
\begin{align}
&-\frac{\lambda\sqrt{\eps}}{|\p \Omega|(\cos\theta-1)} +\sqrt{\eps} \frac{\p^2 v_0(\theta)}{\p \theta^2} =\\
&-\ds \int\limits_{1-\sqrt{\eps}}^1\frac{(4r^{2}\eps+O(\eps^{3/2})) }{|re^{i\theta}(1-\sqrt{\eps})-1+O(\eps)|^4} e^{ -v_0(\theta)}\left(1- \sqrt{\eps} v_1(\theta) +O(\eps)\right )\,dr. \nonumber
\end{align}
that is, the BVP \eqref{PDEw2} in the section becomes the ODE (with respect to $\theta$),
\begin{align}\label{PDEw4}
v''_0(\theta) =&\, -\ds \frac{(4\eps+O(\eps^{3/2}))}{|e^{i\theta}(1-\sqrt{\eps})-1+O(\eps)|^4} \exp
\left\{-\ds v_0(\theta) \right\} -\frac{\lambda}{|\p \Omega|(1-\cos\theta)} \hspace{0.5em}, \\
\left.v'_0(\theta)\right|_{ \theta=\pi}=&\,\frac{-\lambda\sqrt{\eps}}{2|\p \Omega|}, \nonumber\\
\left.v'_0(\theta)\right|_{\theta=c\sqrt{\eps}}=&\, 0.\nonumber
\end{align}
The graph of the solution of \eqref{PDEw4} in $\Omega_{\omega}$ and $\Omega$  is shown in Figure \ref{f:fullcusp}. Equation \eqref{PDEw4} {is obtained by averaging over the radial direction and its solution seems to be a good approximation to \eqref{PDEw2}} only for small $\lambda$. A different approach for large $\lambda$ is discussed in the next section.

A regular expansion for
$\lambda\ll1$,
\beq \label{expansion}
v_0(\theta)=w_0(\theta)+\lambda w_1(\theta)+o(\lambda),
\eeq
} gives in \eqref{PDEw4} that $w_0=O(\eps)$ and $w_1$ is the solution of the BVP
\begin{align}\label{PDEw5}
w''_1(\theta) =&\, -\ds \frac{(4\eps+O(\eps^{3/2}))}{|e^{i\theta}(1-\sqrt{\eps})-1+O(\eps)|^4}  -\frac{1}{|\p \Omega|(1-\cos\theta)} \hspace{0.5em}, \\
\left.w'_1(\theta)\right|_{ \theta=\pi}=&\,\frac{-\sqrt{\eps}}{2|\p \Omega|}, \label{PDEw5bc}\\
\left.w'_1(\theta)\right|_{\theta=c\sqrt{\eps}}=&\, 0.\nonumber
\end{align}
Direct integration with respect to $\theta$ gives
\beq\label{w_1}
w_1(\theta) &=& -\frac{2\theta}{\eps\sqrt{\eps}}\arctan\frac{\theta}{\sqrt{\eps}}+\frac{1}{|\p\Omega|}\ln\sin^2\frac{\theta}{2} +
A\theta+B.
\eeq
Equation \eqref{PDEw5bc} gives $A$ as
\beq
A&=&{\frac {\pi }{\eps^{3/2}}}-\frac{4}{3\pi^{3}}-{\frac {\sqrt{\eps}}{2|\p\Omega|}}.
\eeq
The zero Neumann boundary condition cannot be satisfied and a boundary layer appears, leading to the local expansion
%We then fixe the constant $C$ in \eqref{PDEw5} such as $B=0$ in \ref{w_1}.
\beq \label{expansion2}
v_0(\theta)=\lambda w_1(\theta)+o(\lambda).
\eeq
It follows that for $\lambda \ll1$, the solution increases with $\lambda$. It is shown below that it decreases for $\lambda \gg1$, demonstrating that there is at least one maximum in the variable $\lambda$.
%%%%%%%%%%%%%%%%%%%%%%%%%%%%%%%%%%%%%%%%%%%%%%%%%%
\subsection{PNP asymptotics in a charged disk with a charged funnel}
\begin{figure}[http!]
	\center
	\includegraphics[scale=0.8]{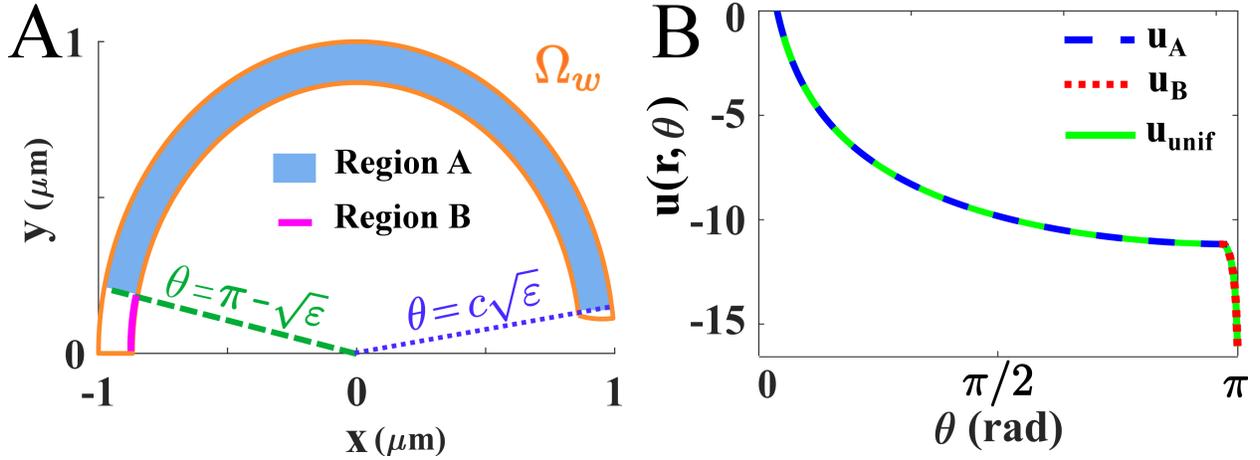}
	\caption{\small\setstretch{1.5} { {Decomposition of the banana-shaped domain $\Omega_w$ into two subregions regions $A$ and $B$.} {\bf A.} Representation of the two subregions
	$A$ (blue) and $B$ (magenta) of $\Omega_w$. {\bf B.} Solutions of
	\eqref{v_regA} (dashed blue), \eqref{uB} (red dots), and the uniform approximation  $u_{unif}$ of  \eqref{unif} (green) for $r=1-\sqrt{\eps}$.}}
	\label{f:RegionsAB}
\end{figure}

%%%%%%%%%%%%%%%%%%%%%%%%%%%%%%%%%%%%%%%%%%%%%%%%%%%
In the limit of $\lambda\gg1$, $\eps\to0$, the asymptotic expansion of the potential found above for a charged disk with a  funnel is no longer valid. Some insight can be gained by observing the field lines in the domain $\Omega_{w}$, described in Figure \ref{f:Field}B. These lines are parallel to the radius vector, except in a small region near the funnel.  Two sections can be distinguished,
\begin{align}\label{regions}
 A=&\{ (r,\theta)\in \Omega_w\,:\,|\theta-\sqrt{\eps}|>\pi,\ |r-1|\leq\sqrt{\eps}\}\nonumber \\
 &\\
 B=& \{w=(1-\sqrt{\eps})e^{i\theta}\,:\,\ |\theta-\pi|\leq\sqrt{\eps}\}. \nonumber
\end{align}
The two sections $A$ and $B$ are illustrated in Figure \ref{f:RegionsAB}A. Note that the boundary of section $B$ contains a circular arc (marked magenta). Next, the approximate solutions $u_{A}(r,\theta)$ and $u_{B}(\theta)$ of \eqref{PDEw2} in the two sections  are constructed
%%%%%%%%%%%%%%%%%%%%%%%%%%%%%%%%%%%%%%%%%%%%%%%%%%%
%%%%%%%%%%%%%%%%%%%%%%%%%%%%%%%%%%%%%%%%%%%%%%%%%%%
and used to construct a uniform approximation $u_{unif}$ in $\Omega_w$ (Fig. \ref{f:RegionsAB}B).

%%%%%%%%%%%%%%%%%%%%%%%%%%%%%%%%%%%%%%%%%%%%%%%%%%%
\subsection{Asymptotics of $u_A(r,\theta)$ in section $A$}
%%%%%%%%%%%%%%%%%%%%%%%%%%%%%%%%%%%%%%%%%%%%%%%%%%%
{The boundary conditions \eqref{bcondi} for the potential equation \eqref{PDEw2} indicate
that the radial derivative is $O(\lambda\sqrt{\eps})\to\infty$
%in the regime $\lambda\eps=O(1)$ as $\lambda\to\infty$ and $\eps\to0$.
Thus the angular derivatives are negligible relative to the radial ones. It follows in a regular expansion of the solution that the $\theta$ derivatives can be
neglected relative to the $r$ derivative and the equation is then solved along the
rays  $\theta=const=\theta_0$ for $r\in[1-\sqrt{\eps},1]$. Thus, to leading order in $\lambda\sqrt{\eps}$,
\begin{align}\label{C1}
u''_{A}(r,\theta_0)+\frac{1}{r} u'_{A}(r,\theta_0)=&\,\frac{-4\eps\exp(-u_{A}) }{|re^{i\theta_0}(1-\sqrt{\eps})-1|^4}\mbox{ for }r\in[1-\sqrt{\eps},1]\\
\left. u'_{A}(r,\theta_0)\right|_{r=1-\sqrt{\eps}}=&\,0 \nonumber\\
\left. u'_{A}(r,\theta_0)\right|_{r=1}=&\,\frac{-\lambda\sqrt{\eps}}{|\p \Omega|(1-\cos\theta_0)}.\nonumber
\end{align}
For $\eps\ll1$, we get} $ |re^{i\theta_0}(1-\sqrt{\eps})-1|^4=|e^{i\theta_0}-1|^4+O(\sqrt{\eps})$. {Setting}
\beq\label{h0}
h(\theta_0)&=&\frac{4\eps }{|e^{i\theta_0}-1|^4},
\eeq
{and}
\beq\label{Translate_A}
v_{A, \theta_0}(r)=-u_A(r,\theta_0)+\ln{h(\theta_0)},
\eeq
{we get
\begin{align}\label{C2}
v_{A, \theta_0}''(r)+\frac{1}{r} v_{A, \theta_0}'(r)=&\exp(v_{A, \theta_0})\\
\left. v_{A, \theta_0}'(r)\right|_{r=1-\sqrt{\eps}}=& 0\nonumber\\
\left. v_{A, \theta_0}'(r)\right|_{r=1}=&\frac{\lambda\sqrt{\eps}}
{|\p \Omega|(1-\cos\theta_0)} .\nonumber
\end{align}
The general solution of \eqref{C2} is given by \cite{PhysD2016}
\beq\label{sol_va}
v_{A,\theta_0}(r)=\ln\frac{C_2^2}{2r^2}-\ln\cos^2\frac{C_2}
{2}(\ln{r}-C_1),
\eeq
where the constants $C_1$ and $C_2$ are determined from the boundary conditions \eqref{C2}. Using
\beq\label{derivative}
v_{A, \theta_0}'(r)&=&\frac{C_2}{r}\tan \frac{C_2}{2}(\ln{r}-C_1)-\frac{2}{r},
\eeq
we find the constant $C_1$ from \eqref{derivative} and from the boundary condition \eqref{C2} at the point $r=1-\sqrt{\eps}$, getting
\begin{align}\label{K_1}
C_1=-\left( \frac{2}{C_2}\arctan\frac{2}{C_2}+\sqrt{\eps}\right)+O(\eps).
\end{align}
This gives in \eqref{derivative} at $r=1$ the transcendental equation for $C_2$,
\beq\label{K2}
C_2\tan \frac{-C_2C_1}{2}=\frac{\lambda\sqrt{\eps}}{|\p \Omega|(1-\cos\theta_0)}+2,
\eeq
hence
\beq\label{limit_pi}
\lim\limits_{\lambda\to\infty}-\frac{C_2C_1}{2}=\frac{\pi}{2}.
\eeq
Now, it follows from \eqref{K_1} that
\beq\label{C1C2}
-\frac{C_2C_1}{2}=\arctan\frac{2}{C_2} +\frac{C_2}{2}\sqrt{\eps}.
\eeq
Note that $\lim_{\lambda\rightarrow \infty}C_2\neq0$, because otherwise we would get the
asymptotic expansion
\beq\label{limC1C2}
-\frac{C_2C_1}{2}=\frac{\pi}{2}+\frac{C_2}{2}(\sqrt{\eps}-1)+O(C^3_2),
\eeq
which leads to
\beq\label{K22}
C_2\tan \frac{-C_2C_1}{2}=\frac{2}{1-\sqrt{\eps}}+O(C_2^2)
\eeq
and contradicts the condition \eqref{K2} in the limit $\lambda\to\infty$.

Then \eqref{limit_pi} and \eqref{C1C2} would imply that
\beq\label{C2_O}
\frac{C_2\sqrt{\eps}}{2}=O(1)
\eeq
and \eqref{C2_O} would give $C_2\gg1$, so that the $\arctan$ term in \eqref{C1C2} drops out, and we would be left with
\beq\label{Anz_K2}
-\frac{C_2C_1}{2}\sim\frac{C_2}{2}\sqrt{\eps},
\eeq
hence
\beq\label{Anz_c1}
{C_1}\sim-\sqrt{\eps}.
\eeq
Expanding the left hand side of \eqref{K2}, using \eqref{limit_pi} and \eqref{Anz_K2}, we obtain that
\beq\label{K2_expand}
\tan\frac{C_2\sqrt{\eps}}{2}=- \frac{2}{C_2\sqrt{\eps} -\pi}+O \left( \frac{C_2\sqrt{\eps}}{2} -\pi /2 \right).
\eeq
Together with \eqref{K2_expand}, the solution of \eqref{K2} is
\beq\label{C22}
C_2\sim\frac{\lambda\pi\sqrt{\eps}}{2|\p\Omega|(1-\cos\theta_0)+\lambda\eps}.
\eeq

With the values of $C_1$ and $C_2$ computed in \eqref{K_1} and \eqref{C22}, the solution $v_{A, \theta_0}$ of \eqref{sol_va} is given
by
\begin{align}\label{u_bar}
v_{A,\theta_0}(r)=&\ln\frac{\eps}{2r^2}\left(\frac{\lambda\pi}{2|\p\Omega|
(1-\cos\theta_0)+\lambda\eps}\right)^2\\
&{-\ln\cos^2\frac{\frac{\lambda\pi}{2}\sqrt{\eps}\left[\ln{r}+\sqrt{\eps}\right]}
{2|\p\Omega|(1-\cos\theta_0)+\lambda\eps}}.\nonumber
\end{align}
Finally, using \eqref{Translate_A} and \eqref{u_bar}, we obtain for ($r,\theta)\in A$,
\begin{align}\label{v_regA}
u_A(r,\theta)=&-\ln\frac{|e^{i\theta}-1|^4}{8r^2}\left(\frac{\lambda\pi}{2|\p\Omega|
(1-\cos\theta)+\lambda\eps}\right)^2\\
&+\ln\cos^2\frac{\frac{\lambda\pi}{2}\sqrt{\eps}\left[\ln {r}+\sqrt{\eps}\right]}{2|\p\Omega|
(1-\cos\theta)+\lambda\eps}.\nonumber
\end{align}
The asymptotic solution $u_A$ is plotted in Figure \ref{f:RegionsAB}B (blue dashed line). Comparison
with numerical solutions for various values of $\lambda$ and $\eps$ is shown in Figure
\ref{f:fullcusp} below.
%%%%%%%%%%%%%%%%%%%%%%%%%%%%%%%%%%%%%%%%%%%%%%%%%%%%
\subsection{The asymptotics of $u_B$ in section $B$}
%%%%%%%%%%%%%%%%%%%%%%%%%%%%%%%%%%%%%%%%%%%%%%%%%%%
The asymptotic solution $u_A(r,\theta)$ in section $A$ cannot
satisfy the boundary condition \eqref{bcondi}  at $\theta=\pi$. Indeed,
\eqref{v_regA} gives $\p u_A(r,\theta)/\p\theta|_{\theta=\pi}
=0$, while the boundary condition  \eqref{PDEw4} is $\p v/\p \theta|_{\theta=\pi}=
-\lambda\sqrt{\eps}/2|\p\Omega|$, so a boundary layer correction is needed.

The boundary layer $u_B(\theta)$ is {an asymptotic solution of \eqref{PDEw2} in section}  $B$, where the $\theta$ derivatives dominate the radial ones. The right-hand-side of {\eqref{PDEw2}} can be simplified for $\eps\ll1$. \\
For $r=1-\sqrt{\eps}$ the approximation
\beq\label{approx_B}
\frac{-4\eps}{|re^{i\theta}(1-\sqrt{\eps})-1|^4}\sim\frac{-\eps}{4}
\eeq
holds, which  does not depend on $r$ and $\theta$. With this simplification in \eqref{PDEw2}, we rewrite $u_B(\theta)$ as
\beq\label{shift_B}
u_B(\theta)=\tilde u_B(\eta)+C_0,
\eeq
where $C_0$  is an additive constant and $\tilde{u}_B$ is a function of $\eta=\theta-(\pi-\sqrt{\eps})$ and solves the BVP
\begin{align}\label{1D_PNP}
\frac{\p^2 \tilde u_B(\eta)}{\p \eta^2}=&-\exp\left\{-\tilde u_B(\eta)\right\}\\
\left. \tilde u'_B(\eta)\right|_{\eta=\sqrt{\eps}}=&- \frac{\lambda\sqrt{\eps}}{2|\p\Omega|}\nonumber\\
\left. \tilde u'_B(\eta)\right|_{\eta=0}=&0\nonumber.
\end{align}
%We remark that from \eqref{approx_B} and \eqref{shift_B}, we should have $C_0=\ln\left(4/\eps\right)$. However, to connect the solution $u_B$ to the solution $u_A$ in $A$ (eq. \ref{v_regA}), we shall determine $C_0$.
%\eeq
%\beq\label{C1}
%u''_{B}(\theta)&=&\frac{-\eps\exp(-u_{B}) }{4}\\
%\left. u'_{B}(\theta)\right|_{\theta=\pi-\sqrt{\eps}}&=& 0 \nonumber\\
%\left. u'_{B}(\theta)\right|_{\theta=\pi}&=&\frac{-\lambda\sqrt{2\eps}}{|\p \Omega|(1-\cos(\theta))}.\nonumber
%\eeq
The solution of \eqref{1D_PNP} (see \cite{PhysD2016}) is
\beq\label{bl_sol}
\tilde u_{B}(\eta)=\ln\cos^2\sqrt{\frac{\lambda}{2I_{\lambda}}} \eta,
\eeq
where $I_{\lambda}$ is the solution of the transcendental equation
\beq
I_{\lambda}=\frac{2|\p\Omega|^2}{\lambda\eps}\tan^2\sqrt{\frac{\lambda\eps}{2I_{\lambda}}}.
\eeq
We obtain to leading order for $\lambda\gg 1$ that
\beq\label{I_lambda}
I_{\lambda}=\frac{2\lambda\eps}{\pi^2}\left(1+\frac{8|\p\Omega|}{\lambda\eps}
\right)+O\left(\frac{1}{\lambda\eps}\right).
\eeq
It follows from \eqref{I_lambda}, \eqref{bl_sol}, and \eqref{shift_B} that for $\theta \in B$, the asymptotic solution is
\beq\label{uB}
u_B (\theta)=\ln\cos^2\frac{\pi}{2}\sqrt{\frac{(\theta-(\pi-\sqrt{\eps}))^2}{\eps}}
\left(1-\frac{2|\p\Omega|}{\lambda\eps} \right)+C_0
\eeq
{(see \eqref{shift_B}).} It is shown in Figure \ref{f:RegionsAB}B (red dots).
%%%%%%%%%%%%%%%%%%%%%%%%%%%%%%%%%%%%%%%%%%%%%%%%%%%
\subsection{A uniform approximation of $u(r,\theta)$ in  $\Omega_w$}
%%%%%%%%%%%%%%%%%%%%%%%%%%%%%%%%%%%%%%%%%%%%%%%%%%%
{A uniform asymptotic approximation $u_{unif}(r,\theta)$ of the voltage $u(r,\theta)$  in
the entire mapped domain $\Omega_{w}$ can be now constructed by matching the the leading
term $u_A(r,\theta)$, given in \eqref{v_regA} in section $A$, with that of $u_B(\theta)$, given in \eqref{uB} in section $B$.}

These approximations agree at {$\theta=\pi-\sqrt{\eps}$}, so we obtain that
\beq
C_0=u_A(1-\sqrt{\eps},\pi-\sqrt{\eps}).
\eeq
%Now we shall extend the function $u_A$ and $u_B$ to $C_{usp}$. We observe, using \eqref{v_regA}, that solution $u_A$, has a reflecting boundary at $\theta=\pi$,
%\beq\label{reflecting_A}
%\left.\frac{\p u_A(r,\theta)}{\p \theta}\right|_{\theta=\pi}&=&0,
%\eeq
%and thus it does not contribute to the boundary condition imposed at $\theta=\pi$ (see eq. \ref{bcondi}). Consequently, we extend the solution $u_A$ to the region $C_{usp}$.
%However, the function $u_B$ \eqref{uB} has singularities in region $A$, and thus cannot be extended outside the region $B$. We used an Heaviside function to truncate $u_B$. Then, we define the solution $V$ in the region $C_{usp}$ as follows,
Thus
\beq\label{unif21}
{u_{unif}(r,\theta)}=
\begin{cases}
u_A(r,\theta) &\hbox{ for }  \theta \in [0, \pi-\sqrt{\eps}] \\
u_B(\theta) & \hbox{ for }  \theta \in [\pi-\sqrt{\eps},\pi].
\end{cases}
\eeq
%\beq\label{unif2}
%V(r,\theta)&=& u_{B}(\theta) \,H(\theta-\pi+\sqrt{\eps})+u_A(r,\theta),
%\eeq
%where $H(x)$ is the Heaviside function.
% We can remark that the solution $V(r,\theta)$ satisfies all the boundaries conditions from \eqref{reduced_1}, since additionally to \eqref{reflecting_A} we have
%\beq
%\frac{\p u_B(\theta)}{\p r}&=&0.
%\eeq
%
The numerical solution of \eqref{NewPb} in $\Omega_w$ and the {approximation $u_{unif}(r,\theta)$} of \eqref{unif21} are shown Fig. \ref{f:RegionsAB}C.
%
%%%%%%%%%%%%%%%%%%%%%%%%%%%%%%%%%%%%%%%%%%%%%%%%%%%
\subsection{Potential drop in $\Omega_w$}
%%%%%%%%%%%%%%%%%%%%%%%%%%%%%%%%%%%%%%%%%%%%%%%%%%%
The potential drop $\tilde \Delta_{funnel} u$ between the center of mass $C$ and the tip of the funnel $S$, is
\beq
{\Delta_{funnel} u=u(C)-u(S).}
\eeq
Due to the axial symmetry of the domain $\Omega$, the center of mass $C$ is at  $r=1-\sqrt{\eps}$, hence \eqref{unif21} gives
\beq
u(S)=u(1-\sqrt{\eps},\pi)\quad\mbox{and}\quad u(C)=u(1-\sqrt{\eps},c\sqrt{\eps}).
\eeq
Recall that the constant $c$ depends on the domain geometry only, and is defined by the conformal mapping $w$ (see relation \eqref{w}).
%\beq\label{c}
%c&=&\frac{\arg({w((1-\eps/2)-iy_c)})}{\sqrt{\eps}},
%\eeq
%where the $((1-\eps/2),-y_c)$ are the coordinates of the center of mass $C$, in domain $\Omega$.
The potential drop $\tilde \Delta_{Cusp} u$ in the funnel can be decomposed as the sum of difference of potential between the two sections, $A$ and $B$. First, the
approximations are
\beq\label{diff_A_exp}
\tilde\Delta u_A &=&u_A(1-\sqrt{\eps},\pi) -u_A(1-\sqrt{\eps},c\sqrt{\eps}).
\eeq
and
\beq\label{diff_B_exp}
\tilde \Delta u_{B}=u_B (\pi)-u_B (\pi-\sqrt{\eps}),
\eeq
so that
\beq\label{D_unif}
\tilde \Delta_{funnel} {u\sim}\tilde \Delta u_{A}+\tilde \Delta u_{B}.
\eeq
Using \eqref{v_regA} in $A$,  we get
\begin{align}\label{v_regA0}
u_A(1-\sqrt{\eps},\theta_0)=&\,-\ln\frac{|e^{i\theta_0}-1|^4}{8(1-\sqrt{\eps})^2}
\left(\frac{\lambda\pi}{2|\p\Omega|(1-\cos\theta_0)+\lambda\eps}\right)^2\nonumber\\
&\,-\ln\cos^2\frac{\frac{\lambda\pi}{2}\sqrt{\eps}(\ln(1-\sqrt{\eps})+
\sqrt{\eps})}{2|\p\Omega|(1-\cos\theta_0)+\lambda\eps}.
\end{align}
For $\eps\ll1$, we get from \eqref{v_regA0} that
\begin{align}\label{expand_term2}
-\ln\cos^2\frac{\frac{\lambda\pi}{2}\sqrt{\eps}(\ln(1-\sqrt{\eps})+
\sqrt{\eps})}{2|\p\Omega|(1-\cos\theta_0)+\lambda\eps}= O(\eps).
\end{align}
Hence, using \eqref{expand_term2} in \eqref{v_regA0}, we get
\beq\label{v_regA1}
u_{A}(1-\sqrt{\eps},\theta_0)=-\ln\frac{|e^{i\theta_0}-1|^4}{8(1-\sqrt{\eps})^2}
\left(\frac{\lambda\pi}{2|\p\Omega|(1-\cos(\theta_0))+\lambda\eps}\right)^2+O(\eps).
\eeq

The approximate solution  $u_A(S)$ at the tip of the funnel $S$ (south pole at $\theta_0=\pi$) is
\eqref{v_regA1}
\beq\label{solution4pi_v1}
u_A(S)=-\ln\frac{2\lambda^2\pi^2}{(4|\p\Omega|+\lambda\eps)^2} +2\ln(1-\sqrt{\eps})+O(\eps).
\eeq
At the center $C$, where $\theta_0=c\sqrt{\eps}$, equation \eqref{v_regA0} gives for
$\eps\ll1$  the $\theta_0$-dependent terms in  \eqref{v_regA1} as
\beq\label{expan01}
|e^{i\theta_0}-1|^4=c^4\eps^2+O(\eps^3),
\eeq
and
\beq\label{expan02}
 2|\p\Omega|(1-\cos c\sqrt{\eps})+\lambda\eps&=&\eps(|\p\Omega|c^2+\lambda)+O(\eps^2).
\eeq
Using \eqref{expan01} and \eqref{expan02}, the expression \eqref{v_regA0} reduces to
\beq\label{solution4c_v0}
u_A(C)={-\ln \frac{c^4} {8}  \left(\frac{\lambda\pi}{|\p\Omega|c^2+\lambda} \right)^2}+2\ln(1-\sqrt{\eps})+O\left(\eps\right).
\eeq
For $\lambda\gg1$, \eqref{solution4c_v0} becomes
\beq\label{solution4c_v1}
u_A(C)={-\ln \frac{\pi^2c^4}{8}}+2\ln(1-\sqrt{\eps})+O\left(\eps,\frac{1}{\lambda}\right){{red}.}
\eeq
Finally, the approximate potential difference $\tilde{\Delta} u_A$ in \eqref{diff_A_exp}, is the difference between  \eqref{solution4c_v1} and \eqref{solution4pi_v1},
\beq\label{diff_A}
\tilde\Delta u_A =-\ln\frac{2\lambda^2\pi^2}{(4|\p\Omega|+\lambda\eps)^2}+\ln\frac{\pi^2c^4}{8} +O\left(\eps,\frac{1}{\lambda}\right).
\eeq
%\beq\label{diff_A}
%\tilde\Delta u_A &=&-\ln\left( \frac{2\lambda^2\pi^2}{(2|\p\Omega|+\lambda\eps)^2}\right)+4\ln c.
%\eeq
%\beq\label{diff_A}
%\tilde\Delta u_A &=&-\ln\left( \frac{2^6\lambda^2|\p\Omega|^2}{(4|\p\Omega|+\lambda\eps)^2}\right)+4\ln c.
%\eeq
For $\lambda\gg1$ \eqref{diff_A} becomes to leading order
%\beq\label{Diff_A_large}
%\tilde \Delta u_A &=&-\ln\left( \frac{2\pi^2}{c^4\eps^2}\right)+O\left( \frac{1}{\lambda}\right) \sim C^{ste}.
%\eeq
\beq\label{Diff_A_large}
\tilde \Delta u_A\sim{-\ln\frac{2^4}{c^4\eps^2}},
\eeq
which is independent of $\lambda$. \eqref{uB} shows that the approximate potential in section $B$ is
\beq\label{ub_eps}
u_B(\pi-\sqrt{\eps})=C_0
\eeq
and
\beq\label{ub_pi}
u_B(\pi)= {\ln \sin^2\frac{\pi|\p\Omega|}{\lambda\eps}} +C_0.
\eeq
Using \eqref{ub_eps} and \eqref{ub_pi} in \eqref{diff_B_exp}, {we obtain
\beq\label{diff_bl}
\tilde\Delta u_{B}= \ln \sin^2 \frac{\pi|\p\Omega|}{\lambda\eps}.
\eeq
For $\lambda\gg1$, \eqref{diff_bl} shows that $\tilde\Delta u_{B}$ is
\beq\label{diff_B_large}
\tilde\Delta u_{B}=-2\ln\lambda+2\ln\frac{|\p\Omega|\pi}{\eps}+O\left(\frac{1}{\lambda^2}\right).
\eeq
Finally, using \eqref{diff_A}, \eqref{diff_bl} and \eqref{D_unif}, we find that the potential drop
is
\begin{align}\label{unif_diff}
\tilde\Delta u=&\ln\sin^2\frac{\pi|\p\Omega|}{\lambda\eps}-\ln\frac{2
\lambda^2\pi^2}{(4|\p\Omega|+\lambda\eps)^2}+\ln\frac{\pi^2c^4}{8}\nonumber\\
&\,+ O\left(\eps,\frac{1}{\lambda}\right).
\end{align}
Again, using \eqref{Diff_A_large}, \eqref{diff_B_large} and \eqref{D_unif} for $\lambda\gg1$ limit,
we get the approximate potential drop as
\beq\label{diff_unif2}
\tilde\Delta u\sim-\ln\lambda^2+2\ln\frac{\pi c^2|\p\Omega|}{4}+O\left(\frac{1}{\lambda}\right).
\eeq
Equation \eqref{diff_bl} shows that for $\lambda\gg1$, the potential drop in the funnel domain occurs mostly in the region $B$. The expression \eqref{unif_diff} is plotted in Figure \ref{f:RegionsAB}A-D (red) and compared to $\ln\lambda^2+const$ (green) and to a two-dimensional numerical solution. The good agreement confirms the validity of the asymptotic expansion and thus confirming the new asymptotic formulas derived here.}
We conclude with the general formula for a dimensional cusp-shaped funnel where $|\p\Omega|=\frac{|\p\tilde\Omega|}{R_c}$ and $R_c$ is the radius of curvature at the cusp
\beq\label{diff_unif2Rc}
\tilde\Delta u\sim-\ln\lambda^2+2\ln\frac{\pi c^2|\p\tilde\Omega|}{4R_c}+O\left(\frac{1}{\lambda}\right).
\eeq

%%%%%%%%%%%%%%%%%%%%%%%%%%%%%%%%%%%%%%%%%%%%%%%%%%%
\begin{figure}[http!]
	\center
	\includegraphics[scale=0.8]{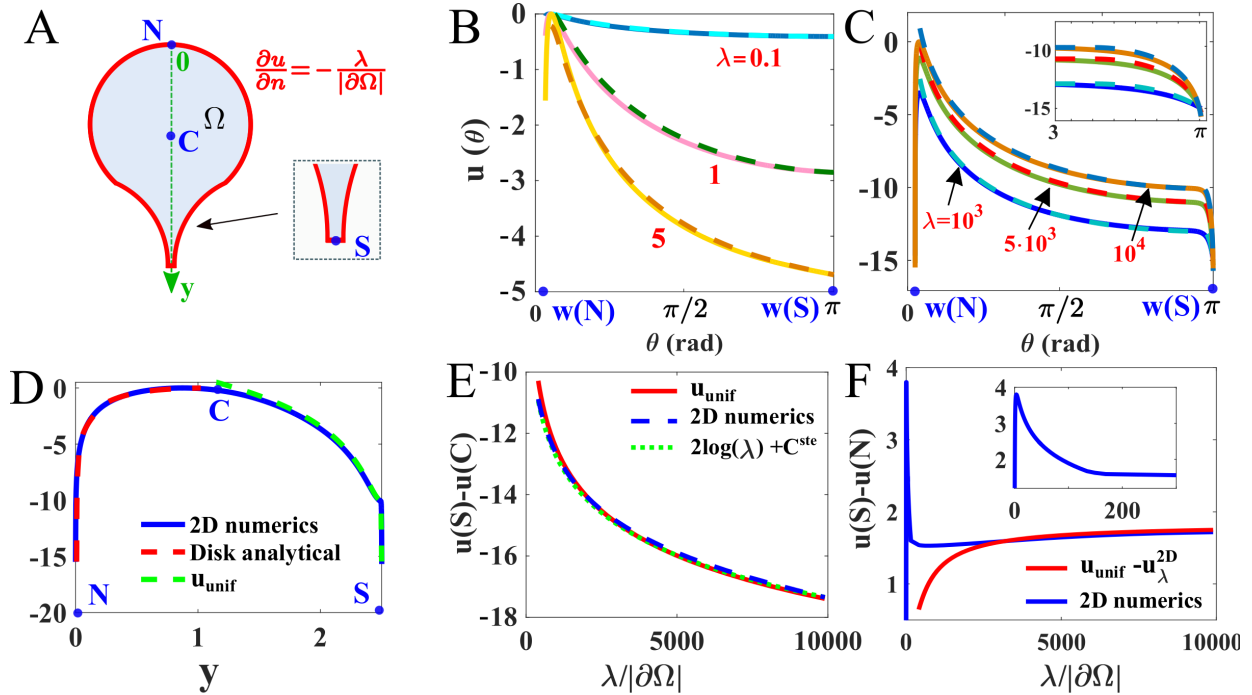}
	\caption{\small\setstretch{1.5} {Comparison of numerical solution of \eqref{NewPb} in the plane with the
approximations $u_{unif}(\x)$ in  \eqref{unif21}}. {\bf A.} Schematic representation of the domain $\Omega$ with a charged funnel (red). The letters $N$, $S$, and $C$ refer to the north pole, the funnel tip, and the center of mass, respectively.
{\bf B-C} Numerical solutions of \eqref{NewPb} (solid) and the solution of  \eqref{unif21} in the funnel (dashed) in the mapped domain  $\Omega_w$ for several values of $\lambda$ and for $\eps=0.01$. {\bf D.} Comparison of \eqref{NewPb} (blue) with analytical solutions \eqref{unif} inside the funnel (dashed green) and \eqref{Solution_Disc} in the bulk (dashed red). {\bf E.} Solution $u(S)-u(C)$ obtained numerically (dashed blue) from \eqref{diffpp3} and analytically from \eqref{unif} (red), compared to the logarithmic function $-2\ln\lambda+const$ (green dots). {\bf F.} Two-dimensional numerical solutions of the difference $|u(N)-u(C)|$ vs $\lambda$ compared to the analytical solutions \eqref{unif_diff} (red). The inset in panel {\bf F.} is a magnification showing a maximum for small $\lambda$.}
	\label{f:fullcusp}
\end{figure}
%%%%%%%%%%%%%%%%%%%%%%%%%%%%%%%%%%%%%%%%%%%%%%%%%%%
%%%%%%%%%%%%%%%%%%%%%%%%%%%%%%%%%%%%%%%%%%%%%%%%%%%%%%%%%%%%%%%%%%%%%%%%%%
\subsection{{Expansion of the potential drop between $N$ and $S$}}
%%%%%%%%%%%%%%%%%%%%%%%%%%%%%%%%%%%%%%%%%%%%%%%%%%%%%%%%%%%%%%%%%%%%%%%%%%%
{To expand the potential difference $u(N)-u(S)$ between the funnel tip $S$ and the north pole $N$ of $\Omega$, we first use
the results \eqref{diff_unif2Rc} computed above, to expand the difference $u(C)-u(S)$, and then subtract \eqref{diff_unif2Rc} and
\eqref{diffpp2dr1}. The} the terms $2\ln(\lambda)$ drop out and we have
\beq
{u(N)-u(S)=2\ln\frac{4|\p\Omega|}{R}-2\ln\frac{\pi c^2|\p\Omega|}{4R_c}+O\left(\frac{1}{\lambda}\right),}
\eeq
where $R$ is the distance between the north pole $N$ and the center of mass $C$ and $R_c$ is the radius of curvature at the cusp.
We obtain to leading order
\beq
{u(N)-u(S)\sim-2\ln\frac{\pi c^2R }{16R_c},}
\eeq
which is a constant that depend only on the center of mass $C$.
%%%%%%%%%%%%%%%%%%%%%%%%%%%%%%%%%%%%%%%%%%%%%%%%%%%
\begin{figure}[http!]
	\center
	\includegraphics[scale=0.8]{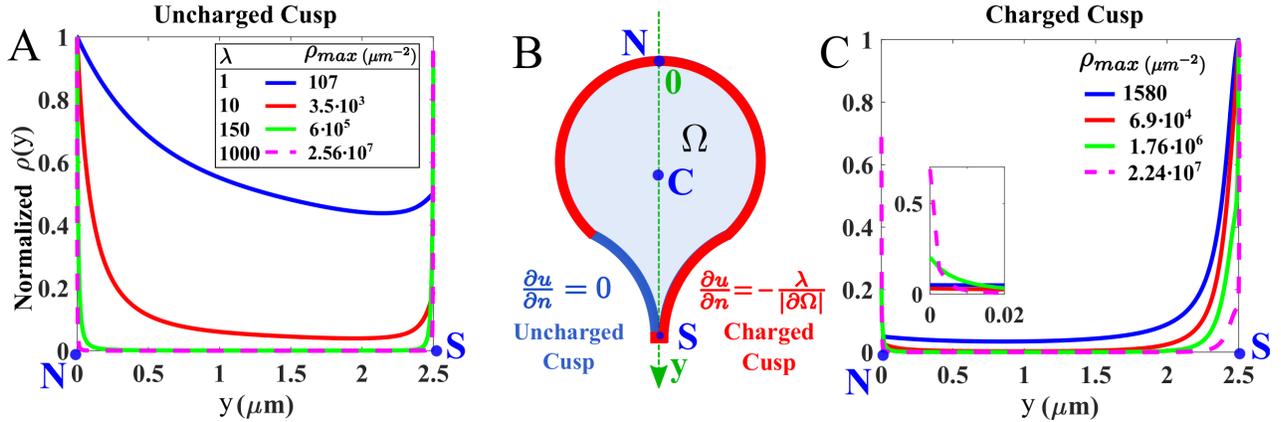}
	\caption{\small \setstretch{1.5}{Normalized charge distribution $\rho(y)$ in charged and uncharged funnel domains. {\bf A.} $\rho(y)$ is computed numerically from \eqref{N} with $\p u/\p n=0$ at the funnel boundaries ($\lambda=1$ (blue), $\lambda=10$ (red), $\lambda=1500$ (green), and $\lambda=1000$ (dashed magenta)). {\bf B.} Representation of $\Omega$ and the funnel
boundary conditions. Left: uncharged funnel domain $\p u/\p n=0$ (blue), and Right: charged funnel domain $\p u/\p n=-\lambda/|\Omega|$ (red). {\bf C.}  $\rho(y)$ in a charged funnel domain.} The same color code is used as in panel {\bf A.}}\label{f:Charges}
\end{figure}
%%%%%%%%%%%%%%%%%%%%%%%%%%%%%%%%%%%%%%%%%%%%%%%%%%%

 %%%%%%%%%%%%%%%%%%%%%%%%%%%%%%%%%%%%%%%%%%%%%%%%%%%%%%
\section{Discussion and conclusion}
%%%%%%%%%%%%%%%%%%%%%%%%%%%%%%%%%%%%%%%%%%%%%%%%%%%%%%
We have derived here new electrostatic laws {in non-neutral confined electrolytes from nonlinear electro-diffusion theory (PNP equations). The effect of local geometrical structure,} such as the local curvature of the boundary {emerges from the asymptotic solution of the model. The PNP equations describe the charge concentration and electric potential. The new electrical laws are derived in the context of non-electro-neutrality, where we use a single ionic species. The approximation of the steady-state solution in a ball with an attached cusp-shaped funnel on its surface is new and the construction of the asymptotic expansion uses a new boundary layer analysis.

Using asymptotic and numerical solution  of the PNP equation, we found here that for a sufficiently high number of charges, the charge concentration peaks at the end of the funnel in a charged funnel boundary domain, but this is not the case for an uncharged funnel
domain (Fig.\ref{f:Charges}A-C). This effect is clearly the result of the cusp-shaped geometry. The present analysis reveals that the curvature affects the membrane potential. We also find that the voltage increases logarithmically in the total number of excess charges $N$, which is valid for uncharged \eqref{diffpp3} and charged \eqref{PDEw2} cusp-shaped funnel on the boundary. We studied here the voltage changes and electro-diffusion under an excess of positive ions. The voltage difference in the limit $\lambda\rightarrow \infty$ is probably attenuated in a mixed ionic solution, but the electro-neutrality remains broken. Cytoplasmic ions are characterized by the following concentrations Na$^+$ = 148ml, K$^+$ = 10ml and Cl$^-$ = 4ml. There is a clear unbalance toward positive charges, however there are probably molecules of various sizes with negative charges to re-balance the charges. However, the motility of these proteins should be driven by a diffusion coefficient smaller than the one of the ions. This difference of mobility is certainly a key feature in maintaining non-electro-neutrality and then tuning the value of $\lambda$. {However, in a system containing an excess of positive and a small amount of negative charges, we show in Appendix that the limit of the PNP equations in the bulk, when the number of negative charges tends to zero, can be obtained by a regular expansion of the solution. This result shows that the small amount of negative charges does not perturb much the distribution of positive ones.} {{red} Finally, note that we did not consider here nanometer structures, such as ionic channels, where a negative ionic charge can affect the motion of the other ions in the channel pore.}

We conclude that local geometrical properties, such as curvature, can modulate the local voltage in biological cellular electrolytes when electro-neutrality is violated. This result generalizes the case of a ball, where the distribution of charges accumulates on the surface as the total charge increases \cite{PhysD2016}.
%We further study the case of charges in a narrow ellipse and inside a corrugated channel. In all cases, the density of charges accumulates near points of local maximum curvature of the boundary. These results can be used in the design of nano-pipettes and help to understand the local voltage changes inside dendrites and axons with heterogenous local geometry.
%We shall discuss several consequences of local charge accumulation at curvature maximum in biological cells First, we have shown that for a corrugated cylinder
Following a non uniform boundary curvature, we expect that charges will be non-uniformly distributed, leading to a difference of potential across the membrane with charges on its surface. {Since, this difference of potential plays a key role in information processing at synapses, we conclude that the spine geometry, in particular its curvature may impact the coding or decoding of voltage through current \cite{YusteBook}.This effect may as well influence the propagation and genesis of local depolarization  \cite{Rall,Qian,HY2015}. More realistic funnels, with two different curvature radii can be incorporated to the formalism presented by modifying the parameter $\alpha$ \eqref{w} as shown in \cite{HS2012}.} The formalism presented in this paper can be applied beyond physiology, in particular in the design of nanopipettes with an optimal shape \cite{Perry,HS2012} by modulating $\alpha$ \eqref{w} or with a patterned surface \cite{Sparreboom} by changing the surface charge density via $\lambda$ in region $A$ \eqref{C1}.}
%%%%%%%%%%%%%%%%%%%%%%%%%%%%%%%%%%%%%%%%%%%%%%%%%%%%%%
\section{Appendix}
%%%%%%%%%%%%%%%%%%%%%%%%%%%%%%%%%%%%%%%%
%%%%%%%%%%%%%%%%%%%%%%%%%%%%%%%%%%%%%%%%
{{red}
\subsection*{Regular expansion of the PNP solution when there are an excess of positive and a small number of  negative charges}
%%%%%%%%%%%%%%%%%%%%%%%%%%%%%%%%%%%%%%%%%%%%%
We show that for the concentrations of cations and anions found in literature \cite{Hille}, the leading order solution of the electrical potential in the bulk can be obtained by considering positive charges only. We assume that the charge of an electrolyte confined in $\tilde\Omega$ consists of identical $N_p$ positive and $N_m$ negative ions with density $q_p(\x)$ and $q_m(\x)$ such as
\beq\label{total_charge}
N_i&=&\int_{\tilde\Omega}q_i(\tilde\x)\,d\tilde\x, \mbox{ for } \, i \in \{ p\,,\,m\},
\eeq
where $p$ and $m$ are positive and negative species respectively. The total charges in $\tilde\Omega$ is the sum
\beq
Q &=&e( N_p-N_m).
\eeq
The associated charge densities $\rho_p(\x,t)$ and $\rho_m(\x,t)$ satisfy the boundary value problem for the Nernst-Planck equation
\begin{align}
D_i\left[\Delta \rho_i(\tilde\x,t) +\frac{z_ie}{kT} \nabla \left(\rho_i(\tilde\x,t) \nabla \phi(\tilde\x,t)\right)\right]=&\,
\frac{\p\rho_i(\tilde\x,t)}{\p t}\hspace{0.5em}\mbox{for}\ \tilde\x\in\tilde\Omega\label{NPE0}\\
D_i\left[\frac{\p\rho_i(\tilde\x,t)}{\p n}+\frac{z_ie}{kT}\rho_i(\tilde\x,t)\frac{\p\phi(\tilde\x,t)}{\p
	n}\right]=&\,0\hspace{0.5em}\mbox{for}\ \tilde\x\in\p\tilde\Omega \label{noflux0}\\
\rho_i(\tilde\x,0)=&\,q_i(\tilde\x)\hspace{0.5em}\mbox{for}\ \tilde\x\in\tilde\Omega,\label{IC0}
\end{align}
where $z_i$ is the valence and $D_i$ is the diffusion coefficient for the ion specie $i$. The electric potential $\phi(\tilde\x,t)$ in $\tilde\Omega$ is solution of the Neumann problem for the Poisson equation
\begin{align}
\label{poisson0} \Delta \phi(\tilde\x,t) =&\,
-\frac{e}{\eps_r\eps_0}( \rho_p(\tilde\x)- \rho_m(\tilde\x))\hspace{0.5em}\mbox{for}\ \tilde\x\in\tilde\Omega\\
\frac{\p\phi(\x,t)}{\p
	n}=&\,-\tilde\sigma(\tilde \x,t)\hspace{0.5em}\mbox{for}\ \tilde \x\in{\p\tilde\Omega},\label{Boundary_Phi0} \nonumber% -\frac{\sigma(\x,t)}{\eps_r\eps_0}
\end{align}
where $\tilde\sigma(\tilde\x,t)$ is the surface charge density on the boundary $\p\tilde\Omega$.
At steady-state, \eqref{NPE0} gives
\beq\label{Boltzman_dist_init}
\rho_i(\tilde\x)&=& \rho_{i,0}\exp\left(\ds-\frac{z_i e\phi(\tilde\x)}{k_BT} \right) \mbox{ for } i\in\{p\,,\,m\},
\eeq
where $\rho_{i,0} $ is obtained from no-flux boundary condition \eqref{noflux0}, thus
\beq\label{Boltzman_dist}
\rho_i(\tilde\x)&=& \frac{ N_i\exp\left(\ds-\frac{z_i e\phi(\tilde\x)}{k_BT} \right)}{\ds\int_{\tilde\Omega}  \exp\left(\ds-\frac{z_i e\phi(\s)}{k_BT} \right)d\s} \mbox{ for } i\in\{p\,,\,m\}.
\eeq
Using the non-dimensionalized potential $\ds\tilde{u}(\tilde\x)=\frac{e\,\phi(\tilde\x)}{k_BT}$,  equation \eqref{Boltzman_dist_init} becomes
\beq\label{Boltzman_dist2}
\rho_i(\tilde\x)&=& \frac{ N_ie^{\ds -z_i\, \tilde{u}(\tilde\x)}}{ \ds \int_{\tilde\Omega}  e^{\ds-z_i\, \tilde{u}(\s)}d\s} \mbox{ for } i\in\{p\,,\,m\}.
\eeq
Using \eqref{poisson0} and \eqref{Boltzman_dist2}, we obtain
\beq\label{PNP_2charges}
-\Delta \tilde{u}(\tilde\x) &=& \frac{l_BN_pe^{\ds- \tilde{u}(\tilde\x)}}{ \int_{\tilde\Omega} e^{\ds- \tilde{u}(\s)}\, d\s}-\frac{l_B N_me^{\ds\tilde{u}(\tilde\x)}}{ \int_{\tilde\Omega} e^{\ds \tilde{u}(\s)}\, d\s}\, \mbox{ in }\, \tilde{\Omega}\\
\frac{\p u(\tilde\x)}{\p n}&=&-\frac{(N_p-N_m) }{|\p\tilde\Omega| }l_B \, \mbox{ on }\, \p\tilde{\Omega},\nonumber
\eeq
where $l_B$ is the Bjerrum length. Using $\ds \x=\frac{\tilde \x}{R_c}$ and $\tilde u(\tilde x)=u(x)$ where $R_c$ is the cusp curvature radius, \eqref{PNP_2charges} becomes 
\beq\label{PNP_2charges2}
-\Delta  {u}(\x) &=& \frac{l_BN_pe^{\ds-  {u}( \x)}}{R_c \int_{ \Omega} e^{\ds-  {u}(\s)}\, d\s}-\frac{l_B N_me^{\ds  {u}(\x)}}{R_c \int_{ \Omega} e^{\ds  {u}(\s)}\, d\s}\, \mbox{ in }\,  {\Omega}\\
\frac{\p u( \x)}{\p n}&=&-\frac{l_B(N_p-N_m) }{R_c|\p \Omega| } \, \mbox{ on }\, \p {\Omega}.\nonumber
\eeq
The small parameter is $\zeta=\ds\frac{N_m}{N_p}\ll1$ because in the bulk, the concentration of negative charges such as chloride (about $4$ mM) is much smaller than positive ones (potassium and sodium account together roughly for $167$ mM \cite{Hille}). A regular expansion of $u(\x)$ is
\beq\label{expand0}
u(\x)&=&u_0(\x)+\zeta u_1(\x)+\cdots
\eeq
Using \eqref{expand0} in \eqref{PNP_2charges2}, in small $\zeta$ limit, we have
\beq\label{PNP_2charges40}
-\Delta  {u_0}(\x) &=& \frac{l_B N_p e^{\ds-  {u_0}( \x)}}{R_c  \int_{ \Omega} e^{\ds-  {u_0}(\s)}\, d\s}\, \mbox{ in }\,  {\Omega}\\
\frac{\p u_0( \x)}{\p n}&=&-\frac{l_B N_p }{R_c|\p \Omega| }\, \mbox{ on }\, \p {\Omega},\nonumber
\eeq
and in $\Omega${\small
\beq\label{PNP_2charges41}
 \Delta  {u_1}(\x) &=& \frac{l_B N_p }{R_c }  \left(\frac{e^{\ds- u_0(\x)}}{\int_{ \Omega} e^{\ds-  {u_0}(\s)}\, d\s}\left( u_1(\x)-
\frac{\int_{ \Omega} u_1(\s)e^{\ds-  {u_0}(\s)}\, d\s}{\int_{ \Omega} e^{\ds-  {u_0}(\s)}\, d\s}
\right)+
\frac{e^{\ds  u_0(\x)}}{\int_{ \Omega} e^{\ds  {u_0}(\s)}\, d\s}\right) \nonumber\\
\frac{\p u_1( \x)}{\p n}&=&\frac{l_BN_p }{R_c|\p \Omega| }\, \mbox{ on }\, \p {\Omega},
\eeq}
which admit a regular solution. This result shows that the limit of the PNP equation when $\zeta$ tends to zero (small charge limit) gives $v_0(\x)$, and thus we conclude that a small amount of negative charges does not perturb the distribution of the total excess of positive charge in the bulk.}
%%%%%%%%%%%%%%%%%%%%%%%%%%%%%%%%%%%%%%%%
\subsection*{The numerical procedure}
%%%%%%%%%%%%%%%%%%%%%%%%%%%%%%%%%%%%%%%%%%%%%
Numerical solutions were constructed by the Comsol Multiphysics 5.0 (BVP problems), Maple 2015 (Shooting problems) and Matlab R2015 (Conformal mapping). The boundary value problems in 1D, 2D, and 3D were solved by the finite elements method in the  Comsol 'Mathematics' package. We used an adaptive mesh refinement to ensure numerical convergence for large value of the parameter $\lambda$.

We solved the {PDEs by the shooting procedure} for boundary value problems using Runge-Kutta 4 method, as well as solvers
from Maple packages. %The sensibility to initial conditions was a key issue plaguing the computations due to the finite element based method {\bf JEROME I DO NOTUNDERSTAND THIS SENTENCE }.

%%%%%%%%%%%%%% BIBLIOGRAPHY %%%%%%%%%%%%%%%%%%%%
\bibliographystyle{plain}

\end{document}